\documentclass[
superscriptaddress,
preprint, showkeys,
 amsmath,amssymb,
 aps,
]{revtex4-1}
\usepackage{braket}
\usepackage{color}
\usepackage{tikz}
\usepackage{graphicx}
\usepackage{dcolumn}
\usepackage{bm}
\newcommand*\circled[1]{\tikz[baseline=(char.base)]{
		\node[shape=circle,draw,inner sep=.3pt] (char) {#1};}}

\begin{document}


\title{Vector magnetometry using electromagnetically induced transparency with $lin\perp lin$ polarization}
\author{Bankim Chandra Das}
\email{ch.bankimdas@gmail.com}
\affiliation{Saha Institute of Nuclear Physics, HBNI, 1/AF, Bidhannagar,
Kolkata -- 700064, India.}
\author{Arpita Das}
\affiliation{Saha Institute of Nuclear Physics, HBNI, 1/AF, Bidhannagar,
	Kolkata -- 700064, India.}
\author {Dipankar Bhattacharyya}
\affiliation{Department of Physics, Santipur College, Santipur, Nadia, West Bengal, 741404, India.}

\author{Sankar De}
\email{sankar.de@saha.ac.in}
\affiliation{Saha Institute of Nuclear Physics, HBNI, 1/AF, Bidhannagar,
Kolkata -- 700064, India.}

\date{\today}

\begin{abstract}
Vector magnetometry was studied using the electromagnetically induced transparency (EIT) with linear $\perp$ linear ($lin \perp lin$) polarization of the probe and the pump beams in $^{87}Rb$ - $D_2$ transition. The dependence of the EIT on the
direction of the quantization axis and the relative orientation of the polarization of the
applied electric fields was studied experimentally. We have shown that from the relative strengths
of the $\sigma$ and $\pi$ EIT peaks, the direction of the magnetic field can be found. Moreover from the relative
separation between the $\sigma$ and $\pi$ EIT peaks, the strengths of the magnetic field can be calculated.
We have also demonstrated that the EIT peak amplitudes show oscillatory behaviour depending upon the orientation of the laser polarization relative to the magnetic field
direction. Using the positions of the maxima and minima, the direction of the magnetic field can be calculated. To understand the experimental observation, a theoretical study has been done numerically considering all the thirteen Zeeman sub-levels. Apart from the numerical model, a toy model has also been considered to obtain an analytical response of the medium considering the velocity distribution. The dependencies of the magnetic field direction and the polarization direction of the electric fields have been explicitly derived in the analytical model. Further the direction of the magnetic field is calculated using the analytical solution. This study can be helpful in order to make an EIT based atomic vector magnetometer at room temperature.
\end{abstract}

\maketitle

\section{Introduction}
The coherent interaction of the electric fields with the atomic medium which leads to the phenomena like electromagnetically induced transparency (EIT) \cite{kasapi95,arpita_2018}, electromagnetically induced absorption (EIA) \cite{lezama,bankim2019,bae} are highly sensitive to the external magnetic fields. Using this fact, magnetic field with good spatial resolution and high sensitivity can be measured. The technological advancements in the field of optical magnetometery, specifically due to the development of precision optical devices and lasers, has helped us to achieve ultra sensitive magnetometers. The optical magnetometers are now driven by the various applications such as heart and brain imaging \cite{boto2018,budker2007}, prediction of the earthquake, detection of the dark matter \cite{AFACH2018162}, fundamental symmetries of nature \cite{Romalis2011}, etc. The commercially available magnetometer like the superconducting quantum interference device (SQUID) \cite{clarke2004squid} is capable of measuring the ultra sensitive magnetic field. But in contrast to the SQUID, the atomic magnetometer does not require cryogenic cooling which opens up an intrinsic advantage of atomic magnetometers for miniaturization \cite{shah2007}. Recent studies have shown that the color centers are promising candidate for miniature magnetometers \cite{Schloss_2018,Jensen_2014}.  Optical pumping magnetometers (OPM) \cite{alexandrov1992} and spin exchange relaxation-free (SERF) magnetometers \cite{Zhang2019} are the highly sensitive magnetometers till date. But all the commercially available magnetometers are not always preferable since these are sensitive to the magnetic field strength only. But the direction of the magnetic field is also important in some of the applications. Using the coherent optical effects like EIT with longitudinal magnetic field (LMF) and transverse magnetic field (TMF) makes an opportunity for developing EIT based vector magnetometers \cite{Yudin_2010, Cox_2011} which is sensitive to the direction of the magnetic field. The effects of LMF \cite{Lampis2016,Bao2016} or TMF \cite{Dimitrijevi_2008,Margalit_2013} in the systems characterized by EIT  or EIA have been extensively studied. 


In this article we have studied the effects of static longitudinal \cite{Iftiquar_2009} and transverse magnetic fields \cite{Margalit_2013,Ram_2010} in a hyperfine $\Lambda$-type EIT system. We have shown how the EIT resonance is highly sensitive to the magnetic field direction as well as the polarization direction of the applied electric fields. The selection rules of the EIT resonances can be controlled by controlling the polarization component of the laser fields with respect to the quantization axis. Further we have shown, how this experimental technique can be useful for building up a vector magnetometer. For the experiment we have chosen $^{87}Rb$ atoms in $D_2$ transition with $lin \perp lin$ polarization of the probe and the pump beams. With this geometry we can easily separate the probe beam from the pump beam and study the probe transmission with perfect clarity. 

In addition to the experiment, Liouvillie's equation was solved numerically considering all the hyperfine Zeeman sub-levels in order to understand the underlying phenomena of the experimental observation. Apart from the numerical solution, a toy model consists of nine level Zeeman sub-system, has been derived in order to understand the phenomena analytically. Using the analytical behaviour of the observed peaks, we have further calculated the direction of the magnetic field from the relative amplitudes of the EIT peaks. The magnitude of the magnetic field is also calculated considering the relative separation between the EIT peaks. The theoretical formulation in existing literature on atomic magnetometry are based on numerical simulations \cite{arimondo1996v, Hollberg2003,Cox_2011} whereas here we have tried to build up an analytical model in addition to numerical simulations.

We have assumed that the quantization axis is along the total magnetic field direction in order to calculate the polarization components of the applied electric fields \cite{Yudin_2010,Dimitrijevi_2008}. The effective polarization changes due to the introduction of both the longitudinal and the transverse magnetic fields. We have used a rotated co-ordinate system in order to calculate the polarization components of the electric fields \cite{Noh_2010,Lee_1998}. We have also introduced the rotation ($\phi$) of the polarization axis of the electric fields along with the magnetic field direction ($\theta$) in the calculation of the components.

\section{Polarization components in a rotated co-ordinate system}

The dipole selection rules depend on the polarization of light. If the polarization of the light is along the quantization axis, then the $\pi$ transition, i.e $\Delta m=0$ will be excited and if the light polarization is perpendicular to the quantization axis, then the $\sigma$ transition i.e. $\Delta m=\pm 1$ will be excited. So, depending upon the polarization component with respect to the quantization axis direction, we can have different combinations of the resonant light.

\begin{figure}[h]
	\centering
	\includegraphics[scale=.4]{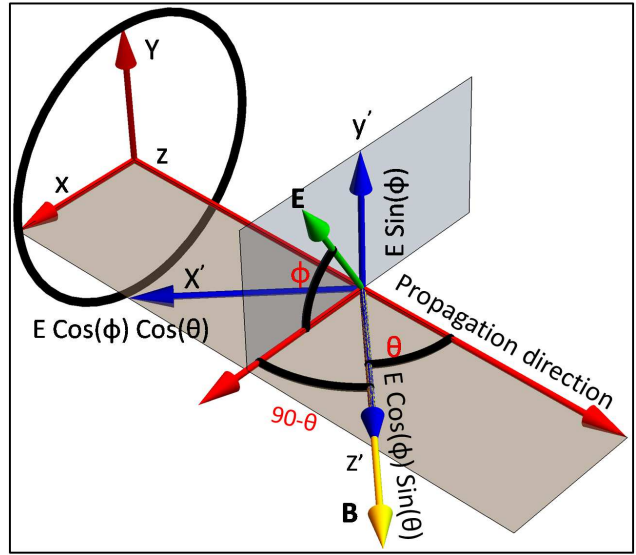}
	\caption{Polarization components in a rotated co-ordinate system $(x',y',z')$.  The magnetic field $\vec{B}$ makes an angle $\theta$ with the $z$ axis. Electric field $\vec{ E}$ makes an angle $\phi$ with the $x$ axis.}
	\label{electric_field}
\end{figure}

Let us consider an electric field with its polarization on the $x-y$ plane and the field is propagating along the $z$ direction (see figure \ref{electric_field}). Let us suppose that a magnetic field is applied on the $x-z$ plane. We will consider the quantization axis to be along the total magnetic field direction. We assume that the polarization vector of the electric field $\vec{E}$ is making an angle $\phi$ with the $x$ axis on the $x-y$ plane and the magnetic field $\vec{B}$ is making an angle $\theta$ with the propagation direction $z$ axis. Now in order to find the probability of the $\pi$ and $\sigma$ transitions, we can simplify our description to calculate the polarization components by introducing an appropriate co-ordinate system $x', y', z'$ where $z'$ is assumed to be parallel to the magnetic field direction and $y'$ is same as the $y$ axis. So the electric field components in the rotated co-ordinate system become $E_{x'}= |\vec{ E}| \cos{(\phi)} \cos{(\theta)}$,
$E_{y'}= |\vec{ E}| \sin{(\phi)}$ and $E_{z'}= |\vec{ E}| \cos{(\phi)} \sin{(\theta)}$.
If $\vec{B}$ is considered to be the quantization axis, then $E_{z'}= |\vec{ E}| \cos{(\phi)}\sin{(\theta)}$ will act as the $\pi$ polarized component because it is parallel to the quantization axis. The other components $E_{y'}= |\vec{ E}|  \sin{(\phi)}$ and  $E_{x'}=  |\vec{ E}| \cos{(\phi)} \cos{(\theta)}$ will act as the $\sigma$ polarized lights since they are perpendicular to the quantization axis. Therefore in this rotated co-ordinate system we can describe the phenomena in a simple way. It is also clear that the intensity of the resonant peak will depend upon the amplitude of the polarization component. Since the polarization components of the electric fields depend on the quantization axis i.e. the direction of the magnetic field, this idea can be used to build up a vector magnetometer. 

\section{experimental methods}

\begin{figure}[h]
	\centering
	\includegraphics[scale=.29]{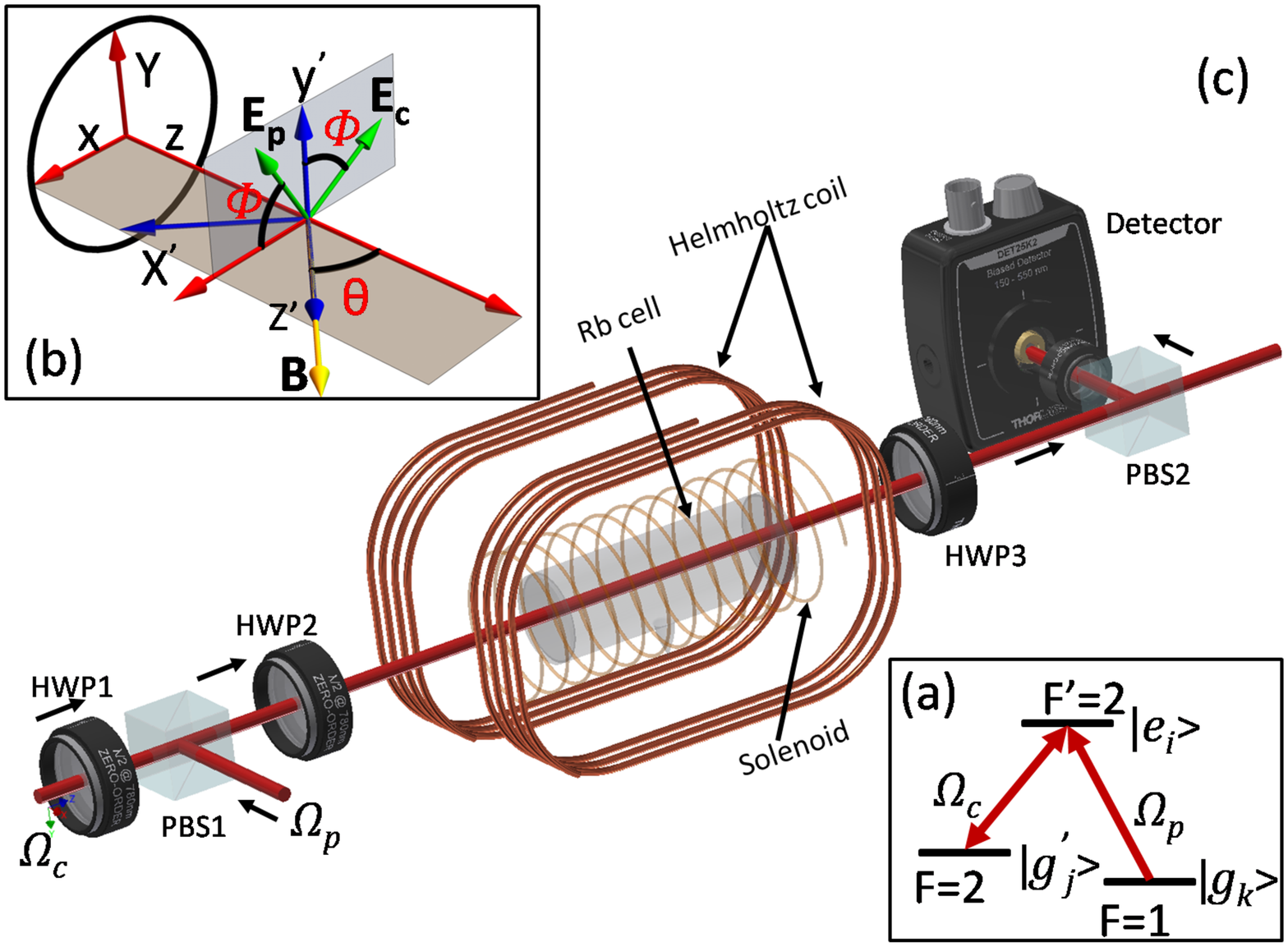}
 	\includegraphics[scale=.29]{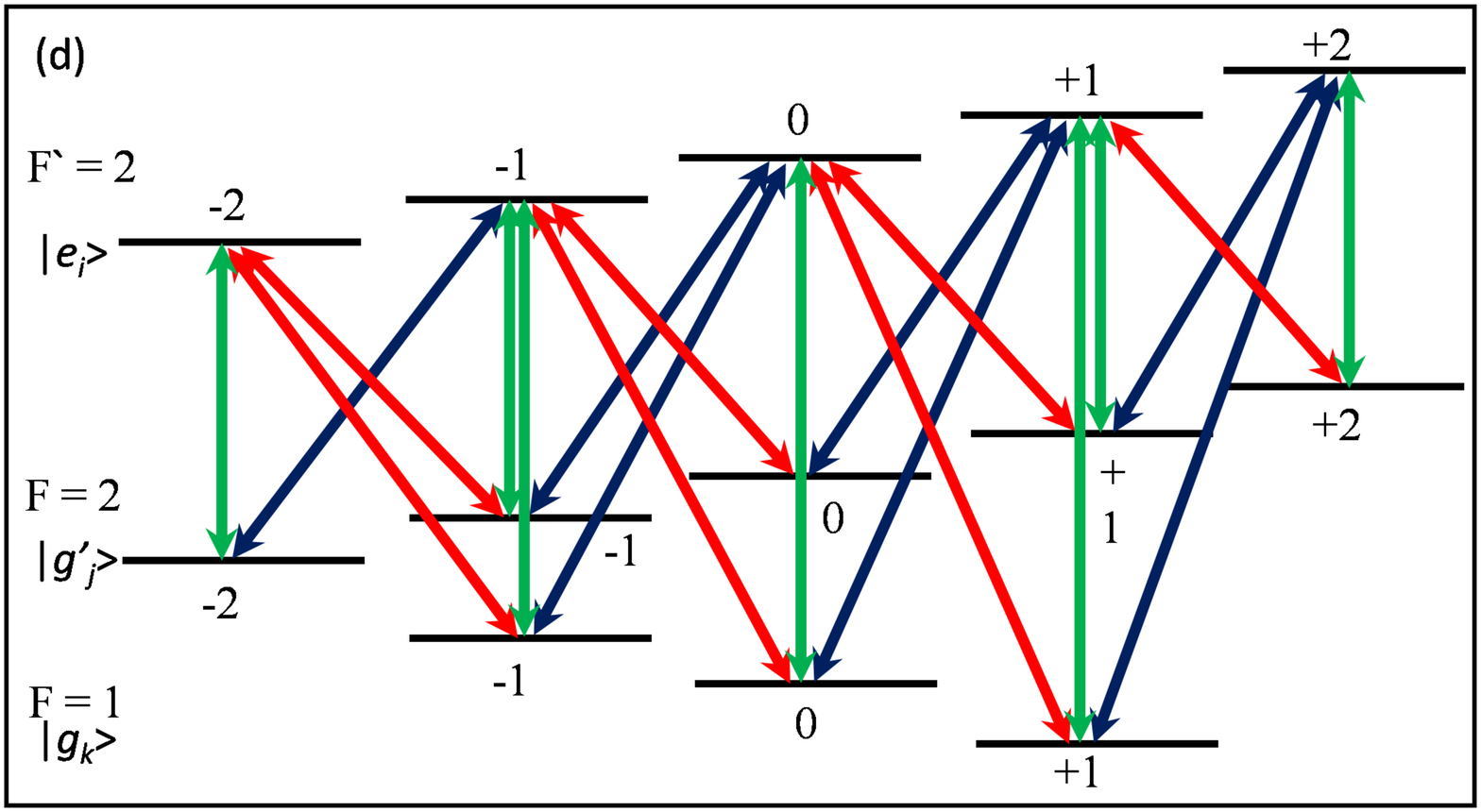}

	\caption{(a) Experimental $\Lambda$-type system of $^{87}Rb$ in $D_2$ transition. (b) Schematic representation of the probe and the pump polarization making an angle $\phi$ with the $x$ and the $y$ axis respectively. $\vec{B}$ is making an angle $\theta$ with the $z$ axis. (c) Experimental setup to study the vector magnetometry. PBS: polarizing cubic beam splitter, HWP: half wave plate, Rb Cell: Rubidium vapour cell. (d) Zeeman sub-levels of the $^{87}Rb$ in $D_2$ transition. $\ket{e_i}$, $\ket{g^{'}_j}$ and $\ket{g_k}$ are assumed to be the excited state, upper and lower ground states respectively for theoretical calculations. The blue, red and the green lines are for $\sigma_{+}$, $\sigma_{-}$ and $\pi$ transitions respectively.
	}
	\label{experimental_diagram}
\end{figure}

In order to study the polarization dependency of the observed peaks experimentally, we have considered a $\Lambda$-type system in $^{87} Rb$-$D_2$ transition (see figure \ref{experimental_diagram}(a)). The polarization of the pump and the probe fields were taken to be linear and mutually orthogonal (see figure \ref{experimental_diagram}(b)). The pump beam ($\Omega_c$) was locked to the transition $F=2 \rightarrow F'= 2$ and the probe beam ($\Omega_p$) was scanned from $F=1 \rightarrow F'= 2$ transition as shown in figure \ref{experimental_diagram}(a). An external magnetic field was applied in an arbitrary direction in the $x-z$ plane with the combination of longitudinal and transverse magnetic fields with respect to the propagation direction as shown in figure \ref{experimental_diagram}(b). The longitudinal field ($\beta_l$) was applied with a solenoid  and the transverse field ($\beta_t$) was applied with a pair of rectangular Helmholtz coils (see figure \ref{experimental_diagram}(c)). The total magnetic field then becomes $|\vec{B}|= \sqrt{\beta_l^2+ \beta_t^2}$. The direction of the magnetic field $\theta =\tan^{-1}(\frac{\beta_t}{\beta_l}) $ was changed by increasing the contribution of one of the magnetic fields. Throughout the interaction region of the Rb vapour cell, the uniformity of the field strength was maintained within $\pm0.01$ $Gauss$. Initially the polarization of the probe beam was parallel to the $x$ axis and the polarization of the pump beam was in the vertical $y$ direction. They were orthogonal to each other. Eventually in our experiment, the mutually perpendicular combination of the probe and the pump beams' polarization was varied with the help of a $\lambda/2$ wave plate (HWP2) with respect to the x axis. After interaction with the medium, the combination was passed through another $\lambda/2$ wave plate (HWP3) in order to make the probe beam polarization parallel to the $x$ axis and the pump beam polarization parallel to the $y$ axis again. Using PBS2, the probe beam was separated from the pump and detected by the detector. The system was kept inside a $\mu$ metal shield in order to reduce the external magnetic field effects (not shown in the figure \ref{experimental_diagram}(c)). We have done the experiment in room temperature in a vapour cell containing Rb atoms in natural isotopic abundance.

After the application of the external magnetic field, the degeneracy of the hyperfine levels was lifted as shown in the figure \ref{experimental_diagram}(d). As mentioned earlier, we have varied the magnetic field direction ($\theta $) and rotated the polarization axis of the electric fields ($\phi$) to study the dependency of the polarization of the electric fields on the probe transmission. For a general case, if the probe and the pump polarization axis makes an angle $\phi$ with the $x$ and the $y$ axis respectively and the magnetic field makes an angle $\theta$ with the $z$-axis, i.e the propagation direction, then according to the previous discussion, the electric field of the probe $\vec{E}_p$ and the pump $\vec{E}_c$ in the rotated coordinate system become, 

	\begin{equation}\label{polarization components}
\begin{array}{ll}
\vec{E}_p&= |\vec{E_p}| \cos{\phi}\cos{\theta} \hat{e_{x'}} +  |\vec{E_p}| \sin{\phi} \hat{e_{y'}} +  |\vec{E_p}| \cos{\phi}\sin{\theta} \hat{e_{z'}}\\
		&= |\vec{E_p}| \cos{\phi}\sin{\theta} \hat{e_{z'}} - \dfrac{|\vec{E_p}|}{\sqrt{2}}(\cos{\phi}\cos{\theta}+ i\sin{\phi} )\sigma_{+}+ \dfrac{|\vec{E_p}|}{\sqrt{2}}(\cos{\phi}\cos{\theta}-i\sin{\phi} )\sigma_{-}\\
\vec{E}_c&= -|\vec{E_c}| \sin{\phi}\cos{\theta} \hat{e_{x'}} + |\vec{E_c}| \cos{\phi} \hat{e_{y'}} -  |\vec{E_c}| \sin{\phi}\sin{\theta} \hat{e_{z'}}\\
		&= |\vec{E_c}| \sin{\phi}\sin{\theta} \hat{e_{z'}}- \dfrac{|\vec{E_c}|}{\sqrt{2}}( -\sin{\phi}\cos{\theta}+ i \cos{\phi}  )\sigma_{+} - \dfrac{|\vec{E_c}|}{\sqrt{2}}( \sin{\phi}\cos{\theta}+ i\cos{\phi}  )\sigma_{-}

\end{array}
\end{equation} 

%
%
These components will act as the $\pi$ and $\sigma$ polarization components with respect to the quantization axis. This situation is schematically depicted in figure \ref{experimental_diagram}(b). Depending on the polarization components, the allowed transitions according to the dipole selection rules are shown in the figure \ref{experimental_diagram}(d). In the experiment we have scanned the probe beam along the transition $F=1 \rightarrow F'=2$ while the pump beam was locked to the transition $F=2 \rightarrow F'=2$. 
While keeping the pump intensity fixed at $17.5$ $mW/cm^2 $ and the probe intensity at $0.1$ $mW/cm^2$, we have studied the effects of the external magnetic field $\vec{B}$ and the dependency of the angle of the polarization axis ($\phi$) of the applied electric fields on the probe transmission. 


\section{Results and Discussions}

%
%

\begin{figure}[h]
	\centering
	\includegraphics[scale=.27]{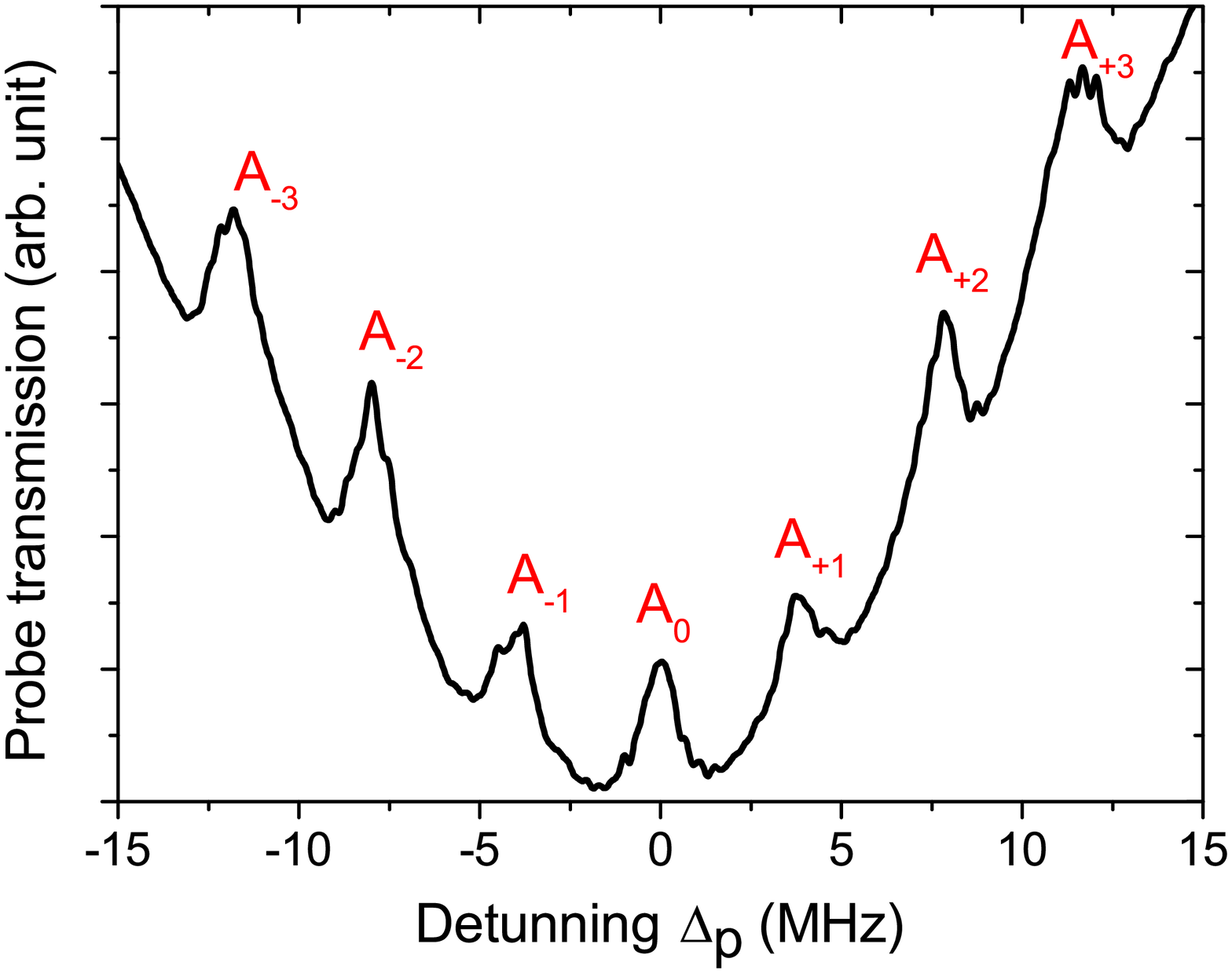}
	\includegraphics[scale=.27]{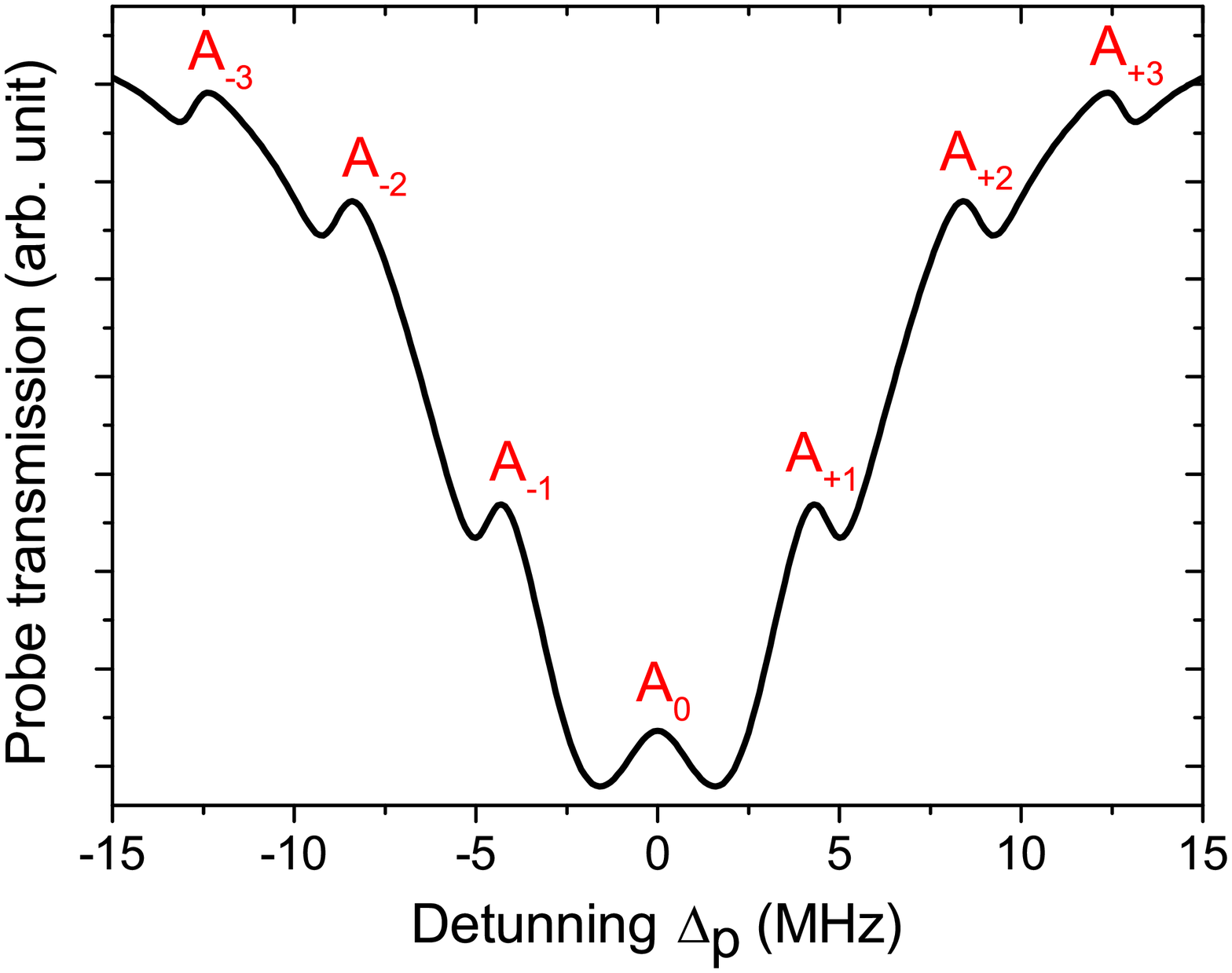}
	\caption{(a) Experimentally measured and (b) numerically solved probe transmission spectra for $\theta= 44.8^0$ and $\phi=40^0$. $A_0$, $A_{\pm2}$ are the $\sigma$ EIT peaks and $A_{\pm1}$, $A_{\pm3}$ are the $\pi$ EIT peaks.
	}
	\label{spectrum}
\end{figure}

In the figure \ref{spectrum}, we have shown how the probe transmission spectra got modified for a general polarization when the quantization axis direction due to the magnetic fields was along $\theta=44.8^0$ and the polarization axis of the probe field was making an angle $\phi=40^0$ with respect to the $x$ axis. In this case the longitudinal field was $\beta_l= 4.26$ $Gauss$ and the transverse field was $\beta_t= 4.23$ $Gauss$.
A total seven EIT peaks have been observed. Here the consecutive peak separation was $\delta= 4.17$ $MHz$, where $\delta= \mu_B g_F |\vec{B}|$ is the energy level shift due to the application of the total external magnetic field $\vec{B}$. $\mu_B$ and $g_F$ are the Bohr magnetron and the Lande-g factor respectively. These EITs are coming due to the contributions of the $\sigma$ and the $\pi$ polarized components of the pump and the probe beams as shown in figure \ref{experimental_diagram}(d). To explain the formation of the observed EIT peaks, we assumed sub $\Lambda$-type systems in the Zeeman levels considering only two photon coherence while we have neglected the higher order coherence. 

%

Since both the pump and the probe beams have both the $\pi$ and the $\sigma$  polarization components, four different combinations of polarization of both the pump and the probe beams may be possible for which the $\Lambda$-type system can be formed. The possible combinations are (a) the pump and the probe both are $\sigma$ polarized light, (b) the probe is $\pi$ polarized and the pump is $\sigma$ polarized (c) the pump is $\pi$ polarized and the probe is $\sigma$ polarized  and (d) both the pump and the probe are $\pi$ polarized light. We will consider each of the combinations separately to explain our experimental results. \\

\begin{figure}[h] 
	\centering
	\includegraphics[scale=.27]{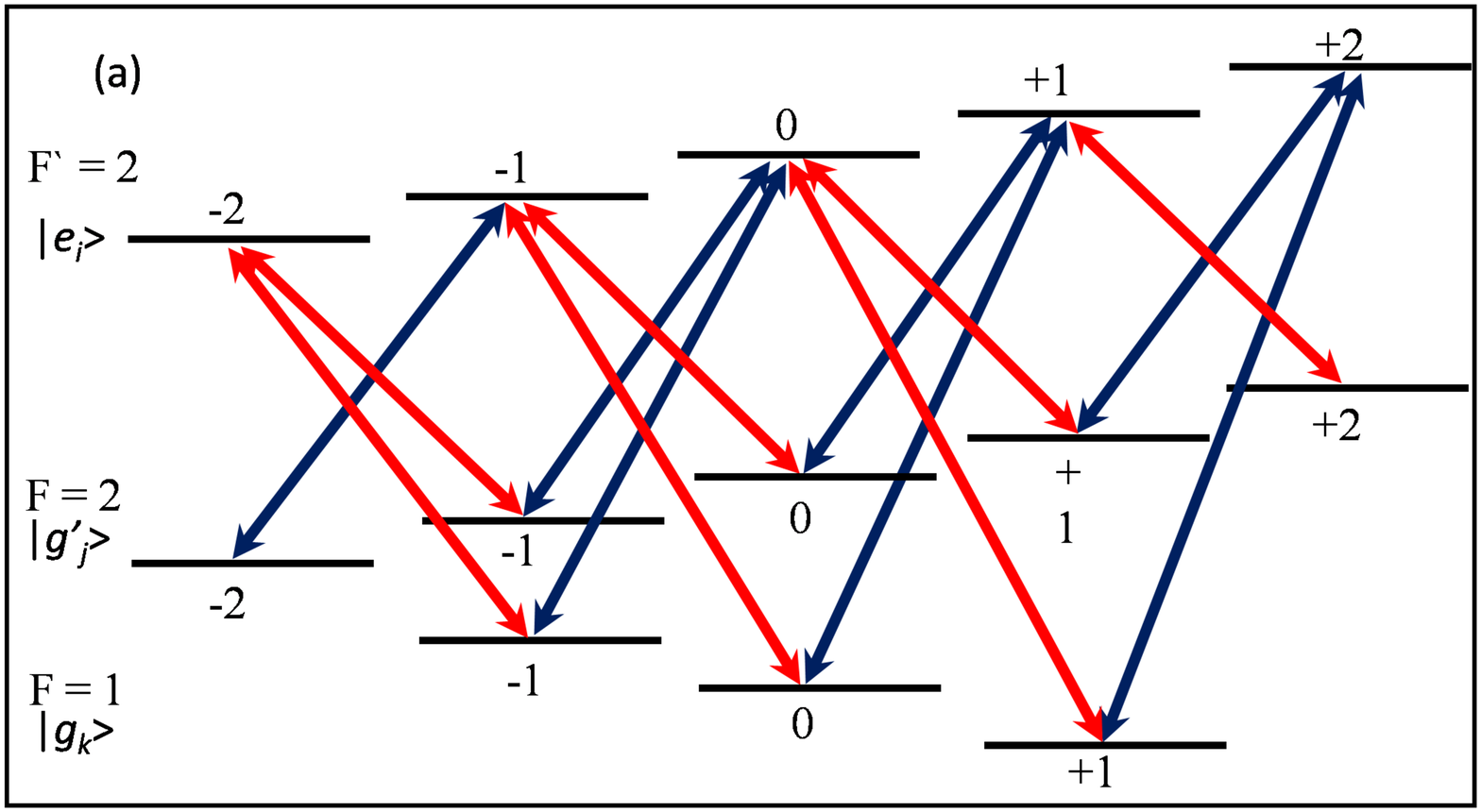}
		\includegraphics[scale=.27]{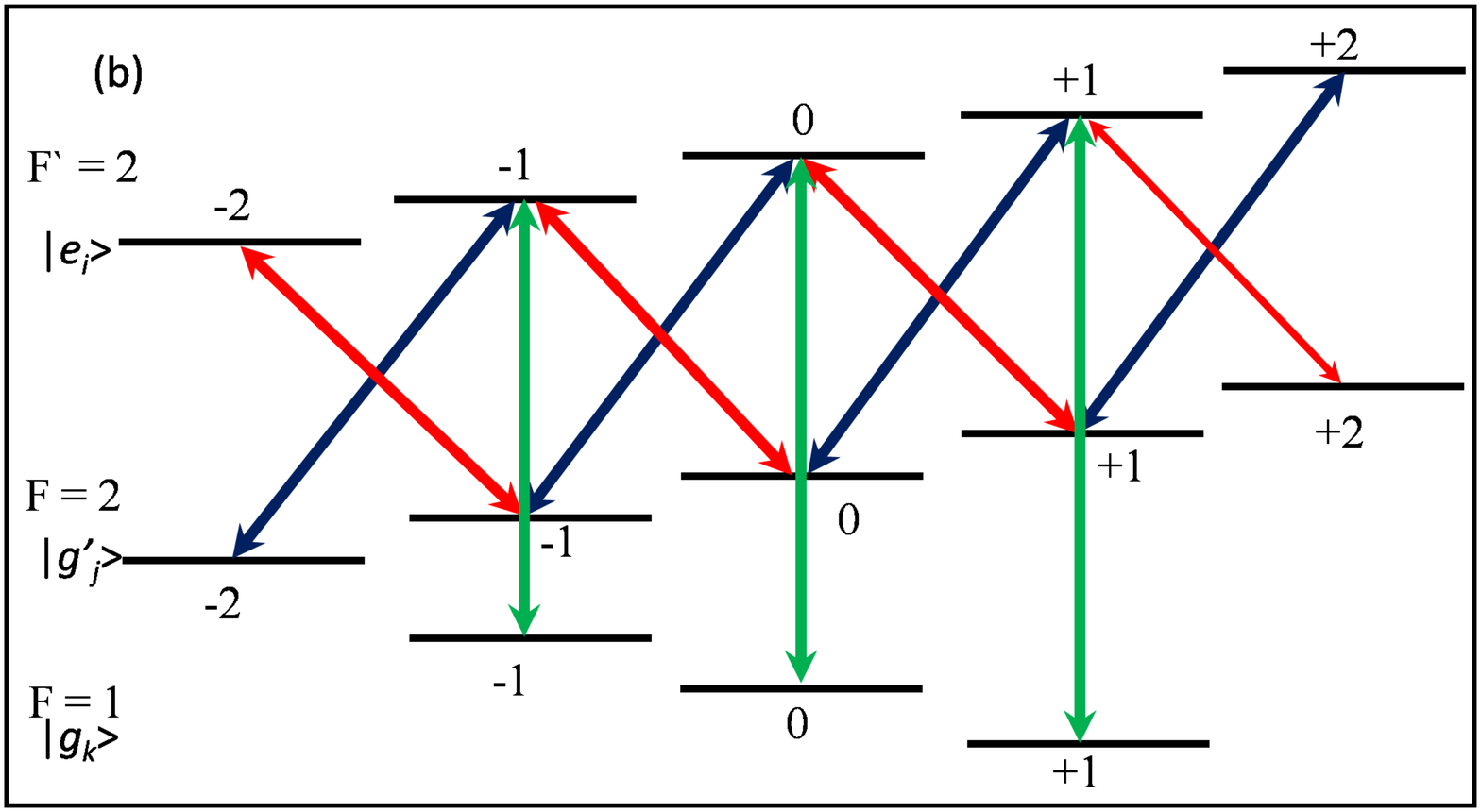}
	\includegraphics[scale=.27]{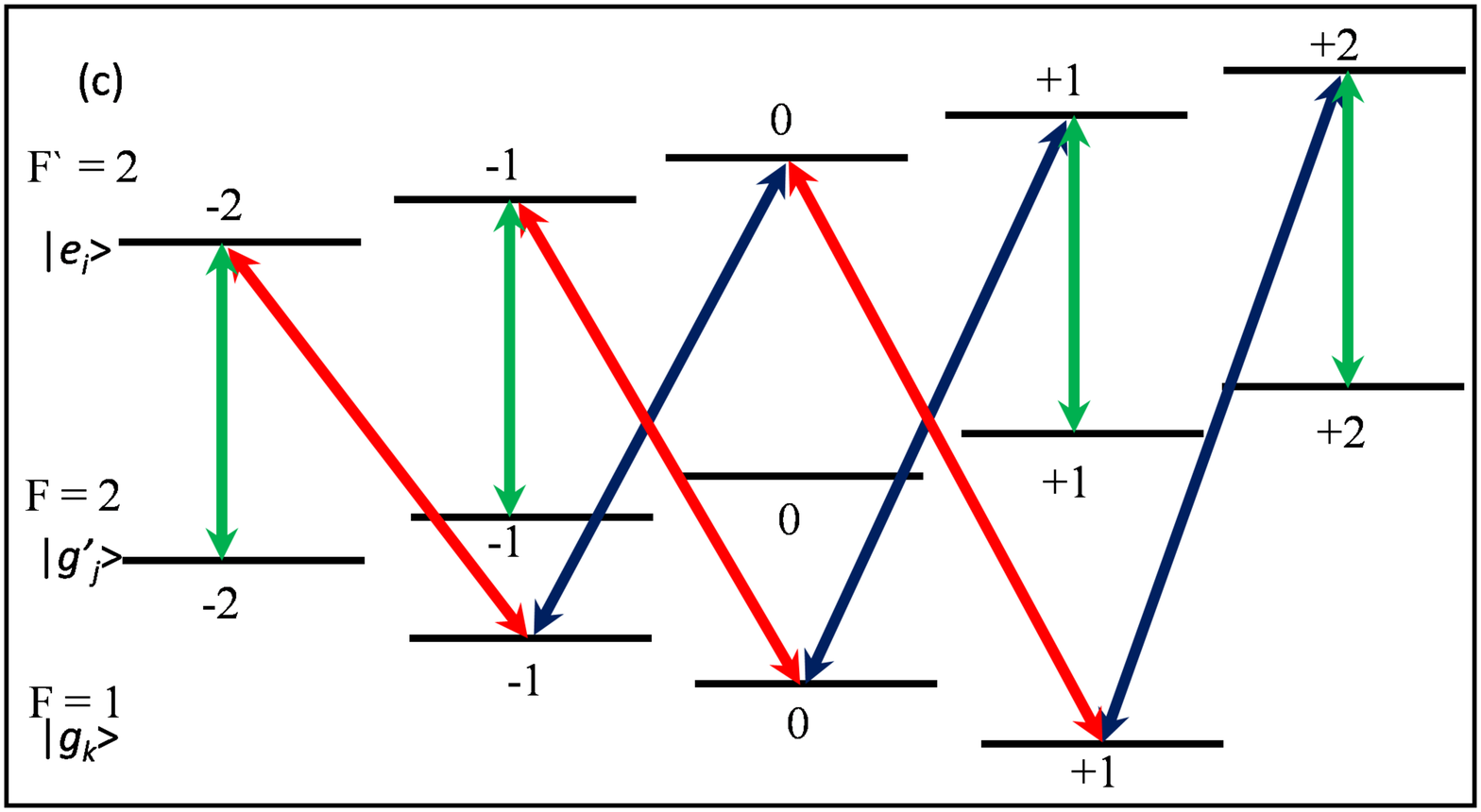}
	\includegraphics[scale=.27]{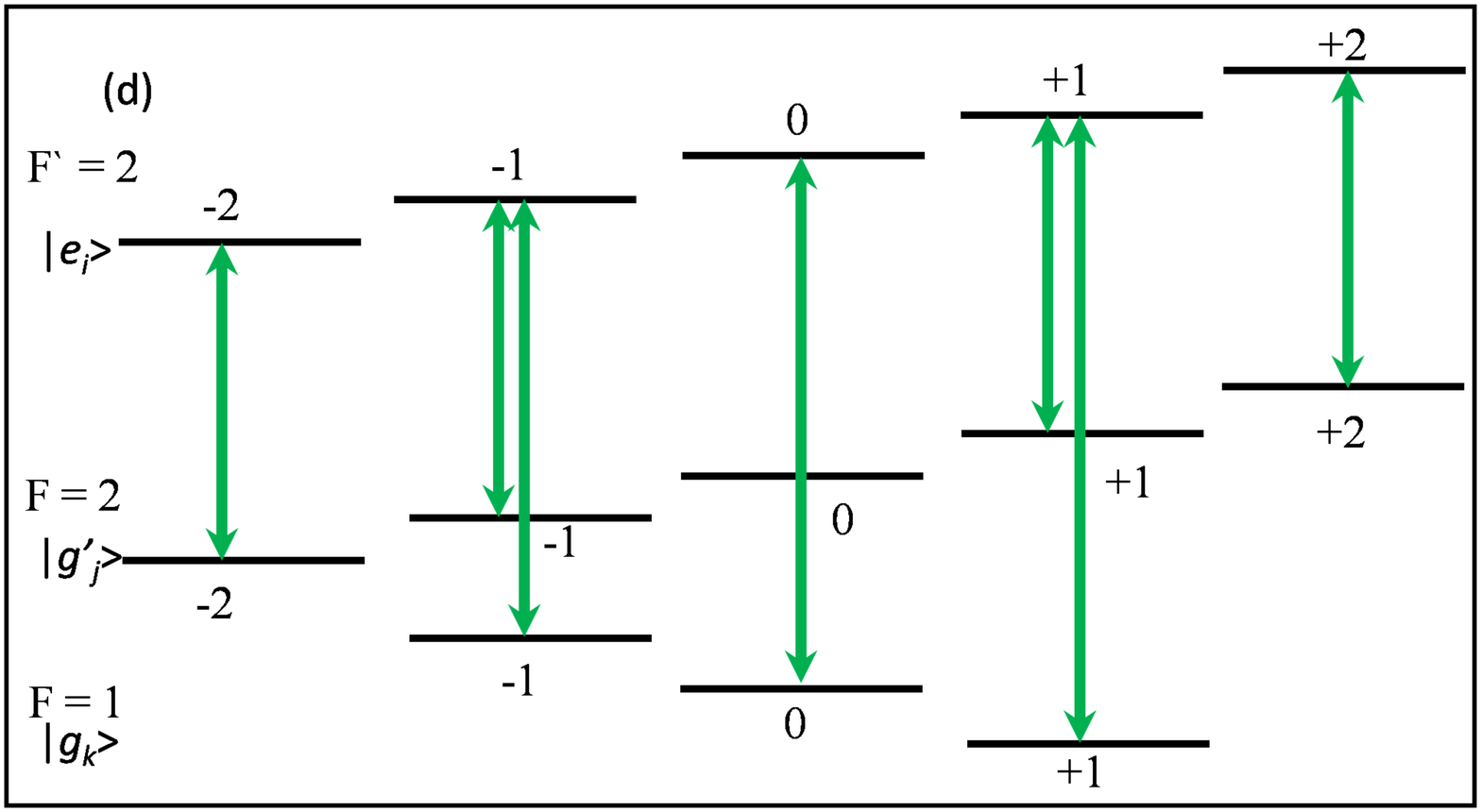}

	\caption{Different combinations of the pump and the probe beams for which the $\Lambda$-type system can be formed. (a) Pump and probe both are $\sigma$ polarized, (b) Probe is $\pi$ polarized and pump is $\sigma$ polarized light, (c) Pump is $\pi$ polarized and probe is $\sigma$ polarized light, (d) Pump and probe both are $\pi$ polarized light.
	}
	\label{sub_lambda}
\end{figure}
Let us assume the first configuration (a), where both the probe and the pump beams are $\sigma$ polarized light.  In this configuration, there can be at the most ten $\Lambda$-type systems possible by the combinations of $\sigma_+$ and $\sigma_-$ components of the probe and the pump beams as shown in figure \ref{sub_lambda}(a). 
The ten EIT resonances will be observed at the positions $0$, $+ 2\delta$ and $-2\delta$ in the probe transmission spectrum depending upon the two photon detunning. So, there will be only three EIT resonance due to the $\sigma$ polarization contributions of the pump and the probe beams. In figure \ref{spectrum}, $A_0$, $A_2$ and $A_{-2}$ have the contributions of this configuration. Specifically, this configuration arises when the magnetic field is perpendicular to the pump and the probe polarizations. If only the longitudinal magnetic field is applied, the probe and the pump beams become $\sigma$ polarized. 

In the figure \ref{sub_lambda}(b) we have considered the second configuration (b) where the probe is $\pi$ polarized and the pump is $\sigma$ polarized. In this case, at the most six  $\Lambda$-type systems can be possible. Here the EIT resonances will come at $\pm \delta$ and $ \pm 3 \delta$ detunnings in the probe transmission depending on the two photon detunning. So, effectively there will be four EIT transmission peaks due to this configuration. In the figure \ref{spectrum}, the peaks $A_{+1}$, $A_{-1}$, $A_{+3}$ and  $A_{-3}$ have contributions of this configuration. This situation can be particularly created  when only the transverse magnetic field is applied i.e., $\theta=90^0$ and the polarization axis of the probe beam is parallel to the $x$ axis while the pump polarization is parallel to the $y$ axis i.e. $\phi =0$. In that case the probe polarization becomes totally $\pi$ but the pump will be $\sigma$ polarized light.

The third possibility (c) is depicted in figure \ref{sub_lambda}(c). Here the probe beam is $\sigma$ polarized light and the pump beam is $\pi$ polarized light. In this configuration, four $\Lambda$-type systems can be formed. The EIT will be observed at the positions $\pm \delta$ and $ \pm 3 \delta$ in the probe transmission similar to the earlier case (b). So, the peaks $A_{+1}$, $A_{-1}$, $A_{+3}$ and  $A_{-3}$ in figure \ref{spectrum} also have the contributions from this configuration. This situation can be created when $\theta=90^0$, i.e., only the transverse magnetic field is applied as in the earlier case but here the polarization axis of the probe beam is perpendicular to the $x$ axis i.e. $\phi =90^0$. Correspondingly the polarization of the pump beam is parallel to the $x$ axis. So, the pump polarization becomes $\pi$ and the probe becomes $\sigma$ polarized light. Interestingly the results for the case (c) is similar to the case (b) but the roles played by the pump and the probe polarizations are completely reversed. 

For the last combination (d) since both the probe and the pump beams are $\pi$ polarized (see figure \ref{sub_lambda}(d)), there will be only two $\Lambda$-type subsystems. The EIT peaks will be observed at $\pm 2 \delta$ in the probe transmission. In the figure \ref{spectrum}, the peaks $A_{\pm 2}$ have contributions from this configuration.
 
 Therefore, when the electric fields have all the $\pi$ and $\sigma$ polarization components, total seven EIT peaks will be observed. The corresponding positions are at $0$, $\pm \delta$, $\pm 2 \delta$ and $\pm 3 \delta$ in the probe transmission. Here three peaks $A_0$ and $ A_{\pm2}$ are due to the longitudinal component of the magnetic field which we termed as $\sigma$ peaks and four EIT peaks $A_{\pm1}$ and $ A_{\pm3}$ are due to the transverse component of the magnetic field which we termed as $\pi$ peaks. In the figure \ref{spectrum}, all the seven EITs have been observed due to all the possible combinations of the pump and the probe beam polarizations as discussed above.
 
Apart from the general case, we have further carried out our study on how the probe transmission was dependent on (a) the magnetic field direction $\theta$ while the angle of the polarization axis $\phi$ was fixed and (b) when the axis of the polarization angle $\phi$ was varied keeping the quantization axis direction $\theta$ fixed. In these cases we have shown that we can control the amplitudes of the $\pi$ and $\sigma$ EIT peaks by changing the polarization components of the pump and the probe fields.

\begin{figure}[h]
	\centering
	\includegraphics[scale=.27]{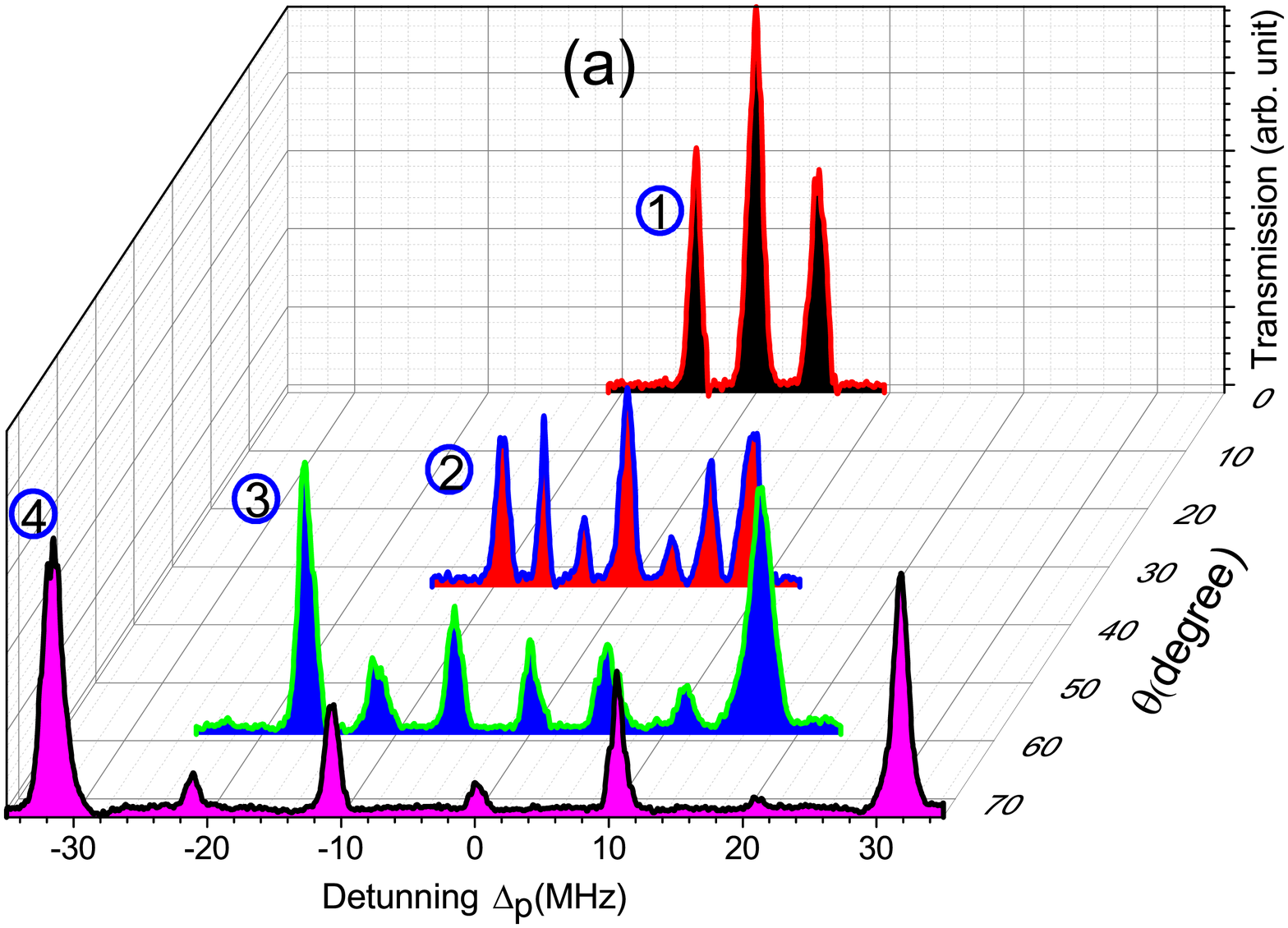}
\includegraphics[scale=.27]{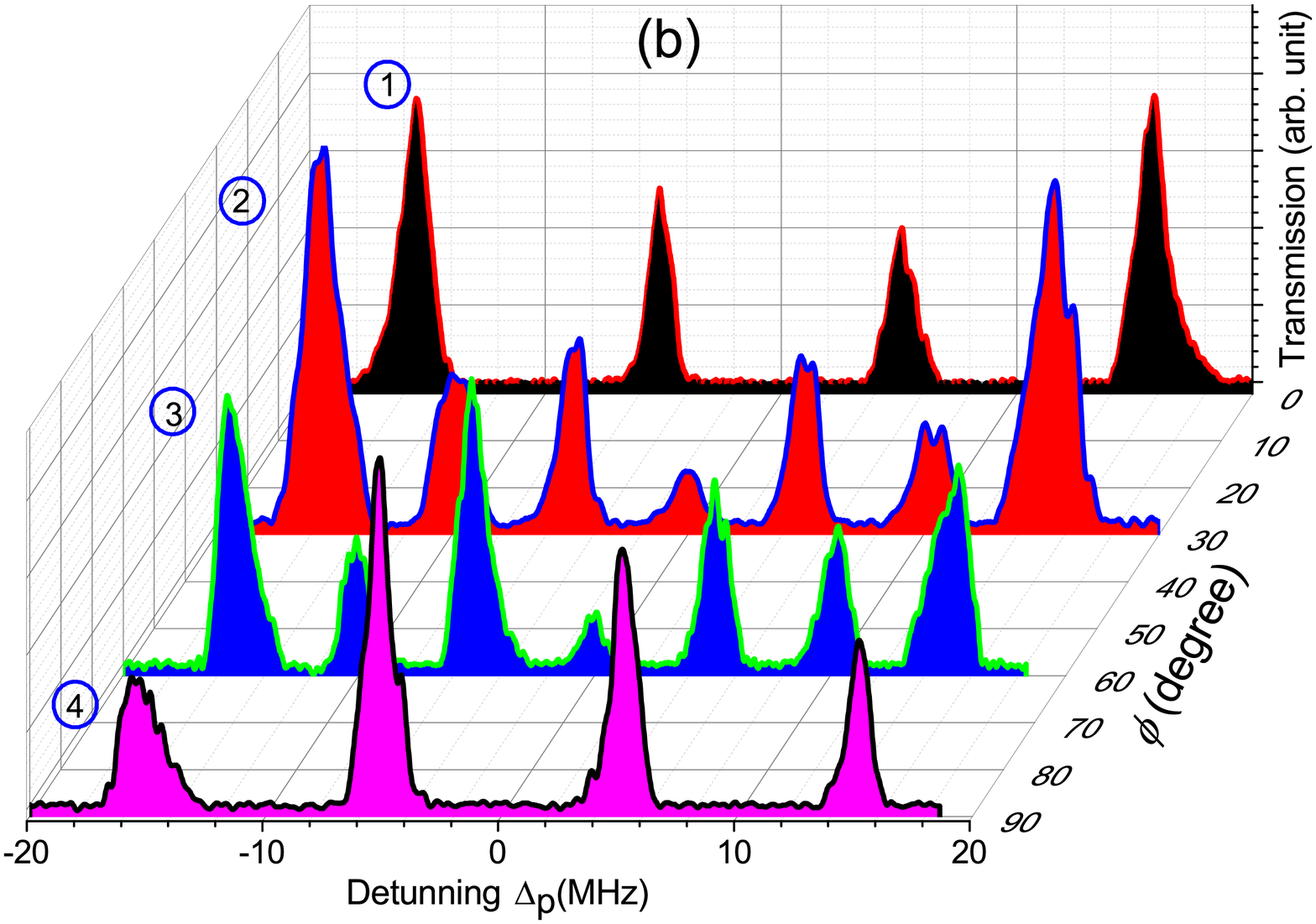}	
	
	\caption{Experimentally observed probe transmission due to the (a) variation of the quantization axis direction $\theta$ for the fixed $\phi =0^0$ and 
		 (b) due to the variation of the polarization axis angle $\phi$ while $\theta$ was fixed at $90^0$. In both the figures, a base line was subtracted from each of the spectra for better visualization of the peak amplitudes. 
	}
	\label{theta_phi_variation_experiment}
\end{figure}

In the the first case (a) we have fixed the angle $\phi=0^0$. So the pump was polarized along the $y$-axis where as the probe beam was polarized along the $x$ axis. Since we have varied the direction of the magnetic field $\theta$, the contributions of the $\pi$ and the $\sigma$ components of the probe electric field will change. In figure \ref{theta_phi_variation_experiment}(a) the observed results have been plotted  for this case. To start with, we have applied only the longitudinal magnetic field ($\beta_l$). The magnitude of $\beta_l $ was kept fixed at $4.26$ $Gauss$. Since in this case $\theta=0^0$, both the pump and the probe beams are $\sigma$ polarized, only three $\sigma$ EIT resonances have been observed (figure \ref{theta_phi_variation_experiment}(a) curve \circled{1}). The coupling scheme is shown in figure \ref{sub_lambda}(a).
Then the transverse magnetic field ($\beta_t$) was introduced and the magnitude was increased gradually. So the direction of the total magnetic field ($\theta$) as well as the total magnetic field strength $|\vec{B}|$ was changed. We have changed the transverse magnetic field from $0$ $Gauss$ to $14.1$ $Gauss$. When the transverse field strength $\beta_t$ is very less compared to the longitudinal field strength $\beta_l$, the contribution of the longitudinal field dominates. In this case the EIT peak strength of $A_0$ and $A_{\pm 2}$ will be higher compared to the other EIT peaks. But when $\beta_t$ is very high compared to $\beta_l$, the EIT peak strength corresponding to the transverse magnetic field i.e $A_{\pm1}$ and $A_{\pm 3}$ will be higher as seen in the figure \ref{theta_phi_variation_experiment}(a) curve \circled{4}. In between these two extreme cases, seven EIT peaks will be observed. So with the increase of $\theta$, the amplitude of the $\pi$ peaks will increase and that of the $\sigma$ peaks will be diminished as observed in figure \ref{theta_phi_variation_experiment}(a). Therefore, in a way we can control the EIT peak strengths depending on the direction of the applied magnetic field.

Since we have increased the magnitude of the total field $|\vec{B}|$, the energy level separation of the Zeeman sub-levels will also increase. Correspondingly the separation of the observed EIT peaks will also increase which is also observed in the figure \ref{theta_phi_variation_experiment}(a). The separation is proportional to the magnitude of the total magnetic field strength. So by measuring the separation between the EIT peaks, we can estimate the magnitude of the total magnetic field $|\vec{B}|$. Further the EIT peak amplitudes are dependent on the direction of the magnetic field. So from the relative amplitudes of the EIT peaks, the direction of the magnetic field ($\theta$) can also be calculated. This method is discussed in section VI.

The observed EIT peak amplitudes are highly dependent on the polarization of the electric fields. In the second case (b) we have directly varied the angle of the polarization axis ($\phi$) while keeping the quantization axis direction fixed at $\theta= 90^0$. In this case we have only applied the transverse magnetic field $\beta_t$. In this configuration, initially when $\phi=0^0$, the probe has only the $\pi$ polarization component and the pump is purely $\sigma$ polarized light. The coupling scheme is similar to the figure \ref{sub_lambda}(b). So, in this case only the $\pi$ EIT peaks ($A_{\pm1}$ and $A_{\pm3}$) will be observed as shown in the figure \ref{theta_phi_variation_experiment}(b) curve \circled{1}. When $\phi$ becomes $90^0$ the situation becomes reverse of the earlier case, i.e.  $\phi=0^0$. The probe polarization becomes purely $\sigma$ and the pump becomes $\pi$ polarized light as in figure \ref{sub_lambda}(c). So, here also we will observe only four $\pi$ EIT resonances ($A_{\pm1}$ and $A_{\pm3}$) as discussed earlier. The experimentally obtained result is plotted in the figure \ref{theta_phi_variation_experiment}(b) curve \circled{4}. In between these two extreme cases when $\phi$ is arbitrary, we will observe the $\sigma$ EIT peaks along with the $\pi$ EIT peaks, since the pump and the probe both will have $\pi$ and $\sigma$ polarization components. As a whole we will observe seven EIT peaks for the other $\phi$ angles (see figure \ref{theta_phi_variation_experiment}(b)). So, it is clear that we can control the strength of the $\sigma$ and the $\pi$ EIT peaks by the angle $\phi$. Also, in this case, the $\sigma$ EIT peaks can be controlled fully from zero intensity upto a maximum value. This can be useful in detecting the magnetic field direction ($\theta$) (see section VI).

\section{theoretical models}
\subsection{Numerical Simulation}
In order to simulate our experimental observations we have solved the density matrix equation considering all the Zeeman sub-levels in the steady state condition keeping $\dfrac{d\rho}{dt}= 0$. In this configuration a total thirteen level system (figure \ref{experimental_diagram}(d)) is formed. Depending upon the selection rules $\Delta m_{F}=0,\pm1$ the $\pi$ and the $\sigma$ polarization components of the light field can couple with the levels. For the theoretical simulation, we assumed $lin \perp lin$ polarization configuration of the pump and the probe beams as considered in the experiment. As mentioned earlier, we assumed that the quantization axis to be along the total magnetic field direction.
The Master equation, after the application of the rotating wave approximations (RWA), becomes 
\begin{equation}\label{optical_bloch}
\dfrac{d\rho}{dt}= -\dfrac{i}{\hbar} \left[\mathcal{H}, \rho \right] - \dfrac{1}{2}\left\{ R, \rho\right\} + \Lambda_A + \Lambda_\gamma
\end{equation}

Here $\mathcal{H}$ is the total Hamiltonian and $\rho$ is the density matrix.
 $R$ is the depopulation matrix or the relaxation matrix which is the sum of the intrinsic relaxation and the transit relaxation rates. The intrinsic relaxation rates depend on the natural decay line width $\Gamma$ and the transit relaxation decay rate {$\gamma$} is proportional to the collisional decay between the ground states. $\Lambda_A$ and $ \Lambda_\gamma$ are the re-population matrices. Here we have considered both the optical re-population and the transit re-population in the re-population matrices.
 Here $\mathcal{H}= H_0+ H_I + H_B$, where $H_0$ is the unperturbed Hamiltonian. $H_0$ is defined as, 
 \begin{equation}
 H_0= \sum_{i} \hbar \omega_{e_i}\ket{e_i}\bra{e_i} +\sum_{j} \hbar \omega_{g^{'}_j}\ket{g^{'}_j}\bra{g^{'}_j} +\sum_{k} \hbar \omega_{g_k}\ket{g_k}\bra{g_k}
 \end{equation}
 The light-atom interaction Hamiltonian $H_I$ is defined as,
 \begin{equation}
 \begin{array}{ll}
H_I&=\sum_{i,j} \ket{e_i}\bra{g_j^{'}} [E_c^{\sigma_+}. \mu_{e_i g^{'}_j} + E_c^{\sigma_-}.\mu_{e_i g^{'}_j} + E_c^{\pi}. \mu_{e_i g^{'}_j}  ]\\
&+ \sum_{i, k} \ket{e_i}\bra{g_k} [ E_p^{\sigma_+}. \mu_{e_i g_k}+ E_p^{\sigma_-}. \mu_{e_i g_k} + E_p^{\pi}. \mu_{e_i g_k}  ] + h.c.
\end{array}
 \end{equation}
The dipole matrix $\mu_{e_i g_k}$ is given by,
\begin{equation}
\mu_{e_i g_k} = \bra{F_e}|er|\ket{F_g} (-1)^{F_e-1+m_g} \left(\begin{matrix}
F_e & 1 &  F_g\\
-m_e & q & m_g
\end{matrix}\right) ;\\ q= 0, \pm 1 
\end{equation} 
In the parenthesis, $3j$ symbol is defined. The magnetic-atom Hamiltonian $H_B$ is defined as, 
 \begin{equation}
 H_B=  \sum_{i}\mu_B {m_{e_i}\ket{e_i}\bra{e_i}} + \sum_{j}\mu_B {m_{g^{'}_j}\ket{g^{'}_j}\bra{g^{'}_j}}+ \sum_{k}\mu_B {m_{g_k}\ket{g_k}\bra{g_k}}
 \end{equation}
Here $\mu_B$ is the Bohr magnetron. $\ket{e_i}$, $\ket{g^{'}_j}$ and $\ket{g_k}$ are the excited state, upper and lower ground states respectively (see figure \ref{experimental_diagram}(d)). $m$ is the magnetic quantum number. $E_{(p/c)}^{\sigma_+}$, $E_{(p/c)}^{\sigma_-}$ and $E_{(p/c)}^{\pi}$ are the electric field amplitudes of the $\sigma_+$, $\sigma_-$, and $\pi$ polarization components respectively of the probe or the pump beams. The probe and the pump amplitudes corresponding to the $\sigma$ and $\pi$ polarizations are dependent on the magnetic field direction $\theta$ and the polarization axis angle $\phi$ as shown in equation \ref{polarization components}. 
All the matrices considered here are $13\times 13$ matrices.
The susceptibility of the medium in the presence of both the transverse and the longitudinal fields becomes,

%
%


\begin{equation}
\chi =\dfrac{N}{\epsilon_0 E_p} \sum_{i, k} \mathbf{\mu}_{e_i g_k}  \rho_{e_ig_k}
\end{equation} 
Here $N$ is the number density of the atoms. Using the above relationship, we can solve $\chi$ for any value of $\theta$ and $\phi$. In order to solve the $\chi$ in the steady state condition, we have solved a set of 169 coupled algebraic optical Bloch equations (equation \ref{optical_bloch}) numerically \cite{ADM}. In figure \ref{spectrum}(b), we have plotted the numerically solved probe transmission considering $\theta=44.8^0$ (corresponding $\beta_l$ and $\beta_t$ values are similar to the experiment) and $\phi=40^0$. All the parameters are taken to be the same as the experimental values. All the seven EIT resonances are observed in the simulated plot in figure \ref{spectrum}(b) since the probe and the pump both have $\sigma$ and $\pi$ polarization components. The simulated peaks in figure \ref{spectrum}(b) are marked according to the experimental spectra as we found them to be coming at the same position as that in figure \ref{spectrum}(a). 
It is clear that in the interaction process, only the two photon contribution is dominating and all the other higher order coherences are negligible. In this way, we can explain all the phenomena after considering the sub $\Lambda$-type system as we had qualitatively described earlier in experimental section IV.\\
Although the numerical solution gives us all the possible EIT peaks and its dependency on the $\theta$ and $\phi$, it is unable to give further insight about the explicit functional dependency of the parameters on the probe transmission. Instead if we are able to form an analytical model, the observed experimental results can be understood in a more efficient way. Since we have shown both qualitatively and numerically that two photon resonance has dominating contribution in the experiment, we can use the series of $\Lambda$-type sub systems in order to make a toy model for an analytical solution. In the analytical solution we have considered the velocity distribution of the atoms whereas in the numerical solution only the zero velocity group of atoms are considered. Further, using the analytical model we can explain how this experiment can be used to build up a vector magnetometer.  
%

\subsection{Analytical Model}

\begin{figure}[h]
	\centering
	\includegraphics[scale=.3]{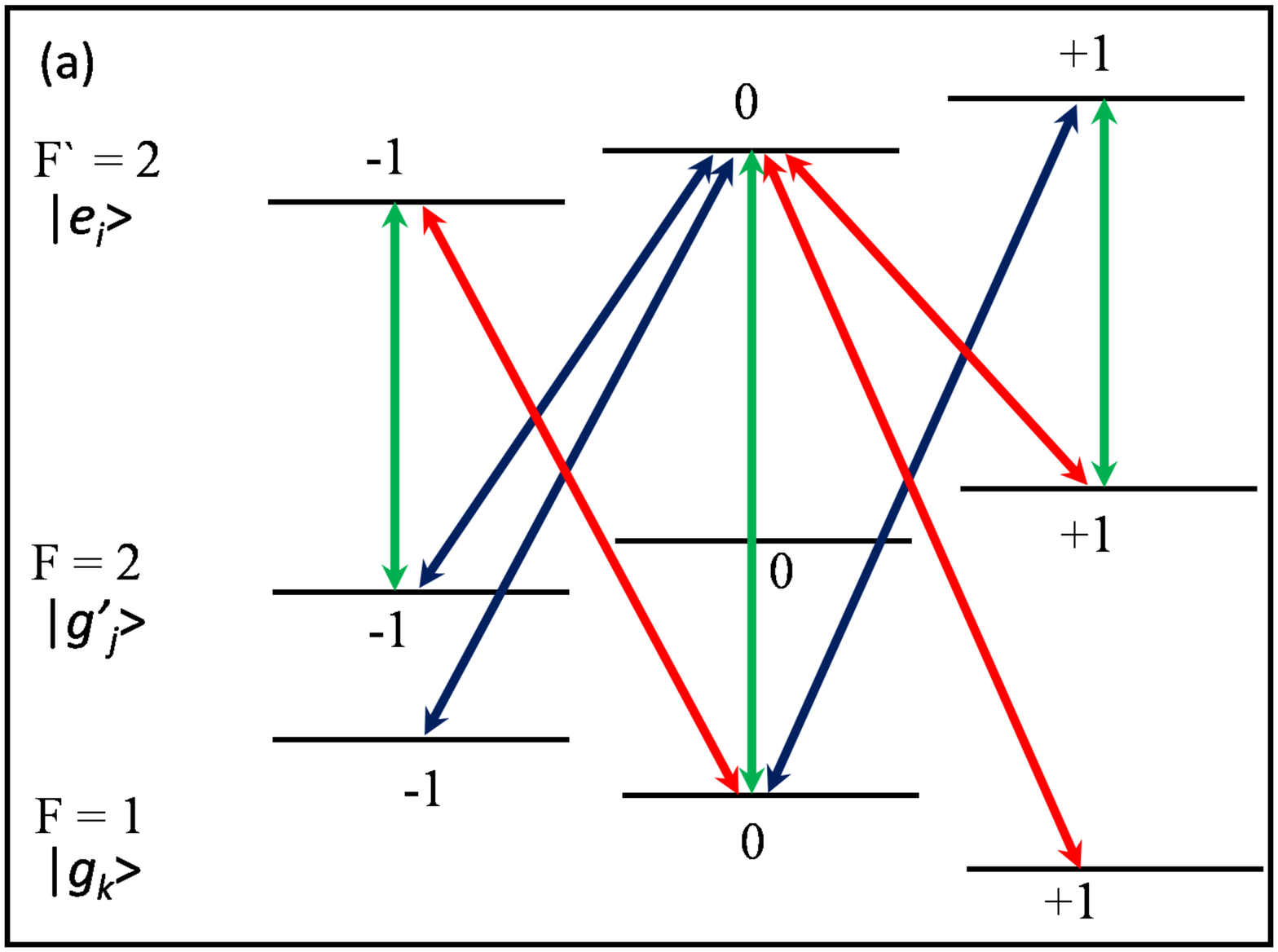}
\includegraphics[scale=.3]{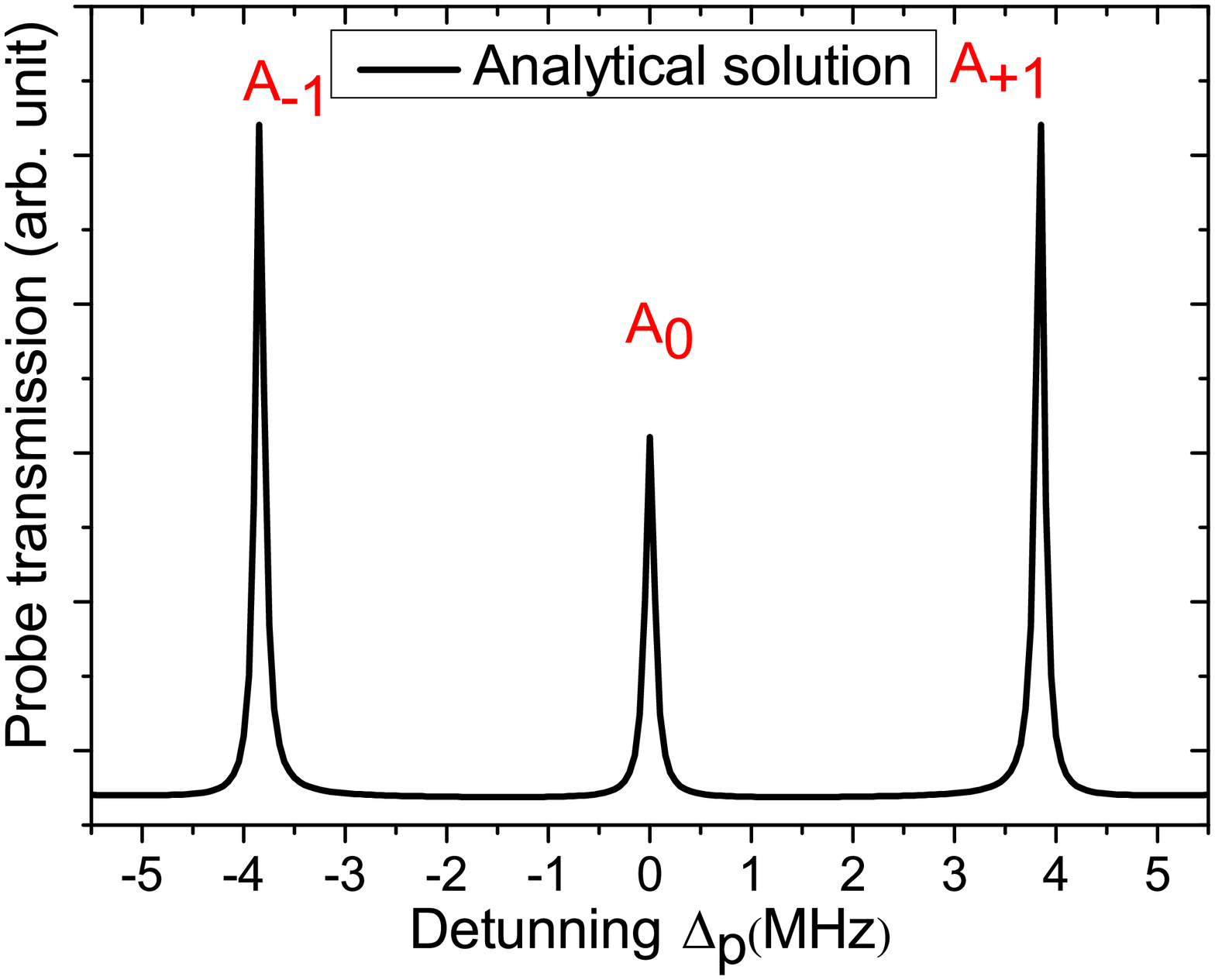}
	
	\caption{ (a) Level diagram of the toy model considering nine Zeeman sub-levels. (b) Analytical solution of the toy model. $A_0$ , $A_{\pm1}$ are the same as the observed peaks shown in figure \ref{spectrum}. Here we assumed $\theta=44.8^0$ and $\phi=40^0$ as in the earlier cases.
	}
	\label{toy_model}

\end{figure}

In order to understand the dynamics in a more quantitative way we have formed a toy model of nine level Zeeman sub-systems to calculate the probe transmission analytically as shown in figure \ref{toy_model}(a). We have considered only one EIT peak that comes due to the contribution of the longitudinal magnetic field namely, $\sigma$ EIT ($A_0$) and two EIT peaks (for symmetry) which comes due to the transverse component of the magnetic field namely, the $\pi$ EIT ($A_{\pm1}$) to keep the model simple.  

The $A_0$ EIT peak arises due to the contribution of the $\Lambda$-type system formed by $\ket{g^{'}_{-1}} \rightarrow \ket{e_{0}}\rightarrow \ket{g_{+1}}$ and $\ket{g^{'}_{+1}} \rightarrow \ket{e_{0}}\rightarrow \ket{g_{-1}}$ transitions. Both the $\Lambda$-type systems have the resonance at $\Delta_p =0$. Here both the pump and the probe beams are $\sigma$ polarized light. Similarly the $A_{+1}$ resonance occurs due to the contribution of the $\Lambda$ type system formed by the transitions $\ket{g^{'}_{-1}} \rightarrow \ket{e_{0}}\rightarrow \ket{g_{0}}$ and $\ket{g^{'}_{-1}} \rightarrow \ket{e_{-1}}\rightarrow \ket{g_{0}}$. Both these $\Lambda$-type systems have resonance at  $\Delta_p =+\delta$. Here one of the pump or the probe beams is $\sigma$ polarized and the corresponding other one is $\pi$ polarized light. 
In the same manner, the $A_{-1}$  peak occurs due to the the contribution of the $\ket{g^{'}_{+1}} \rightarrow \ket{e_{0}}\rightarrow \ket{g_{0}}$ and $\ket{g_{+1}^{'}} \rightarrow \ket{e_{+1}}\rightarrow \ket{g_{0}}$ $\Lambda$-type systems. These two $\Lambda$-type systems have resonance at $\Delta_p =-\delta$. All these configurations are the simplified version of the case represented in the figures \ref{sub_lambda}(a,b and c). In order to include the contribution of both $\theta$ and $\phi$ simultaneously, we have to consider at least two $\Lambda$-type systems for each of the peaks. So, we have considered two $\Lambda$-type system for each of the peaks. Considering all the $\Lambda$-type sub-systems, the total susceptibility $\chi$ of the system including the velocity distribution of the atoms becomes, 
\begin{equation}\label{analytical chi}
\begin{array}{ll}
\chi&= \dfrac{1}{\epsilon_0 E_p}\int_{-\infty}^{\infty} d(kv) N(kv) \sum_{i, k} \mu_{e_i g_k}\rho_{e_i g_k} \\
    &= \dfrac{1}{\epsilon_0 E_p}\int_{-\infty}^{\infty} d(kv) N(kv)\sum_{i, k}\dfrac{i}{6}\dfrac{ \mu_{e_i g_k} \Omega_{e_ig_k}}{\gamma_{e_ig_k}+ i (\Delta_{e_i g_k} +kv)+ \dfrac{|\Omega_{e_ig^{'}_j}|^2}{4\left[\gamma_{g^{'}_jg_k + i \left(\Delta_{e_ig_k}-\Delta_{e_ig^{'}_j}\right)}\right]}}
\end{array}
\end{equation} 
Here we assumed all the populations are trapped in the dark state $\ket{g_k}$, that is $\rho_{g_kg_k}= 1/3$; ($k=0, \pm1$) and $\rho_{g^{'}_jg^{'}_j}=\rho_{e_ie_i} =0$, for the simplicity of the solution. $ \Omega_{e_ig_j}= \frac{\mu_{e_i g_k} E_{e_i g_k}}{\hbar}$ are the Rabi frequencies and $E_{e_i g_k}$ are the electric field amplitudes of the corresponding transitions. $\gamma_{e_ig_{k}}$ are the decay terms between the $\ket{e_i}$ and $\ket{g_k}$. $N(kv)$ is the velocity distribution of the atoms who obey the Maxwell-Boltzmann velocity distribution. But instead of M–B velocity distribution which is a Gaussian distribution, if we assume the distribution to be a Lorentzian having the same FWHM as the Doppler i.e. $2{W}_{D}=2\sqrt{\mathrm{ln}2}{ku}$, the above equation \ref{analytical chi} can be solved analytically \cite{bankim_2018,arpita_2018}. Thus, for the analytical solution, we assumed a Lorentzian velocity distribution of the atoms \cite{javan_2002} defined as,
\begin{equation}
N (k v) = N_{0} \Lambda_0 \dfrac{W_D/\pi}{W_D^2 + (k v)^2}
\end{equation}
Here $N_0$ is the number density of the atoms. $k$ is the wave vector and $\Lambda_0$ is a constant taken to be $\sqrt{\pi \ln 2}$. The equation \ref{analytical chi} can be solved using the contour integral method.
$\chi$ has three poles in the complex $kv$ plane. $kv= \pm i W_D$ and $ kv = i\gamma_{e_ig_{k}}+ i\frac{|\Omega_{e_ig^{'}_j}|^2}{4(\gamma_{g^{'}_{j}g_{k}}+i(\Delta_{e_ig_k}-\Delta_{e_ig^{'}_{j}}))} - \Delta_{e_ig_j}$ for each of the $\rho_{e_ig_k}$. We will consider only the contribution of the pole $kv= - i W_D$. The solution becomes,
 \begin{equation}\label{analytical_solution}
 \chi= \dfrac{i N_{0} \Lambda_0 }{6\epsilon_0 E_p} \sum_{i, k}\dfrac{ \mu_{e_i g_k} \Omega_{e_ig_k}}{\gamma_{e_ig_k}+ i (\Delta_{e_ig_k} -i W_D)+ \dfrac{|\Omega_{e_ig^{'}_j}|^2}{4\left[\gamma_{g^{'}_jg_k + i \left(\Delta_{e_ig_k}-\Delta_{e_ig^{'}_j}\right)}\right]}}
 \end{equation} 
Using this above equation \ref{analytical_solution} we can analytically plot the probe transmission spectrum for any values of $\theta$ and $\phi$. In the figure \ref{toy_model}(b) the probe transmission is plotted assuming $\theta=44.8^0$, $\phi=40^0$ and $\delta=4.17 $ $MHz$ as considered in the experimental data shown in figure \ref{spectrum}. As the $\phi$ and the $\theta$ are arbitrary directions, the pump and the probe both have $\pi$ and $\sigma$ polarization components. So all the possible three EIT ($\pi$ and $\sigma$) peaks are observed. Furthermore, using this formulation, we can explain the specific cases as we have done in the experiment. 
%
%

\begin{figure}[h]
	\centering
	\includegraphics[scale=.27]{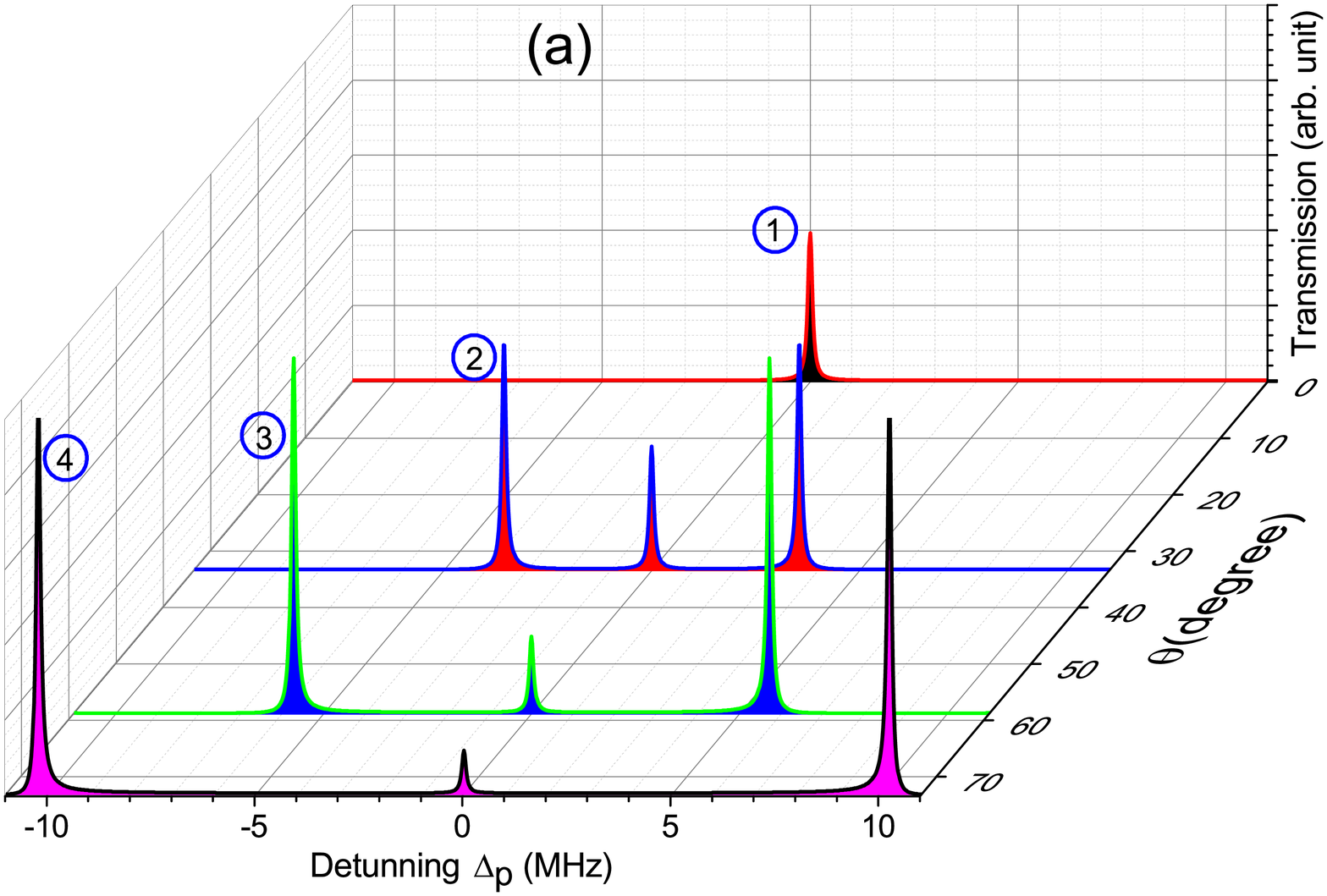}
	\includegraphics[scale=.27]{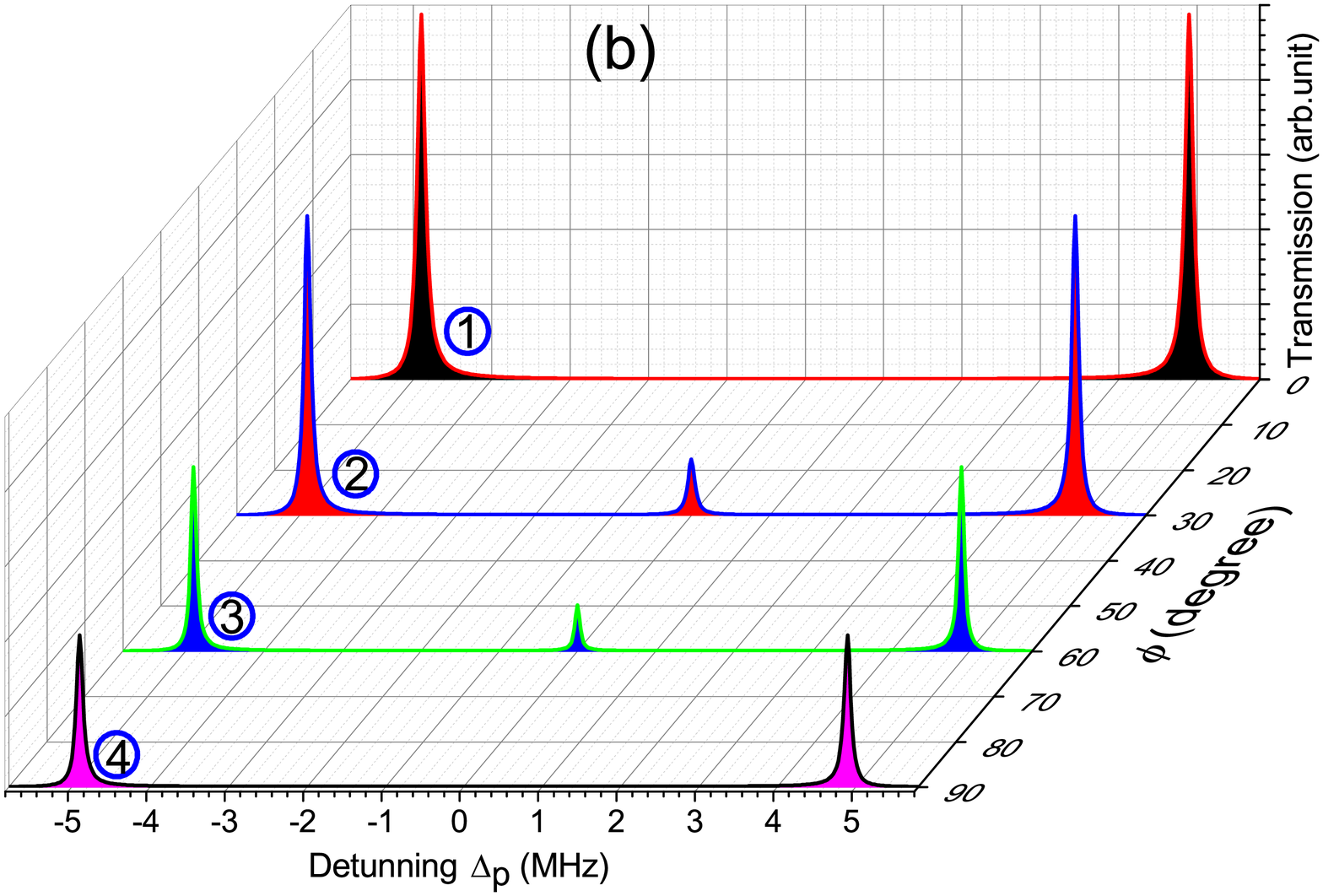}

		\caption{ Analytically plotted probe transmission spectra due to the (a) variation of the quantization axis direction $\theta$ for fixed $\phi =0^0$ and (b) due to the variation of the polarization axis $\phi$ for fixed $\theta= 90^0$. For both the figures, the two photon absorption is subtracted from the one photon Doppler absorption to get the peak amplitudes.}    
	\label{theta_phi_variation_theory}
\end{figure}

For the first case, $\phi=0$ and we have varied the $\theta$ by increasing the $\beta_t$ keeping $\beta_l$ constant at $4.26$ $Gauss$. In the figure \ref{theta_phi_variation_theory}(a) we have plotted the analytical solution of the probe transmission for the variation of $\theta$. Initially when $\theta$ is zero, i.e. the pump and the probe both are fully $\sigma$ polarized light, only the $A_0$ EIT will be observed (figure \ref{theta_phi_variation_theory}(a) curve \circled{1}). Now if we increase the $\theta$ by increasing the contribution of the transverse magnetic field, the peaks $A_{\pm1}$ will appear along with the $A_0$ peak. We will observe all the three resonances until $\theta = 90^0$. In this case the $\sigma$ EIT peak $A_0$ will completely disappear and only the $\pi$ EIT peaks $A_{\pm1}$ will remain. In figure \ref{theta_phi_variation_theory}(a) we have shown the spectra upto $\theta= 74^0$ (curve \circled{4}) similar to the experiment.  Here also we have observed that if the transverse magnetic field strength $(\beta_t)$ is greater than the longitudinal component $\beta_l$, the $\sigma$ EIT peak will be diminished and the other peaks $A_{\pm1}$ will be enhanced as we observed in the experiment (figure \ref{theta_phi_variation_experiment}(a)). Since the transverse magnetic field $\beta_t$ was increased, the separation between the peaks also increased (see figure \ref{theta_phi_variation_theory}(a)).

For the second case, we have plotted the results for the $\phi$ variation keeping the quantization axis direction fixed at $\theta=90^0$. Initially when $\phi=0$, the probe is $\pi$ polarized and the pump is $\sigma$ polarized. It is expected that only the $\pi$ EIT peaks will be observed here. So, in this case two EIT peaks $(A_{\pm1})$ were observed (figure \ref{theta_phi_variation_theory}(b) curve \circled{1}). When $\phi=90^0$ the pump polarization becomes $\pi$ and the probe polarization becomes $\sigma$. So, here also we will observe only two $A_{\pm1}$ EIT peaks (figure \ref{theta_phi_variation_theory}(b) curve \circled{4}). For all the other angles $\phi$, we shall observe the $A_0$ EIT peak along with the $A_{\pm1}$ since the pump and the probe beams both have all the polarization components. Depending on the value of the polarization components, the peak amplitudes will be modified as shown in figure \ref{theta_phi_variation_theory}(b).  Here also we observed similar results as that in the experiment (figure \ref{theta_phi_variation_experiment}(b)). In the above two plots \ref{theta_phi_variation_theory}(a) and \ref{theta_phi_variation_theory}(b) we have subtracted the two photon absorption from one photon Doppler background analytically i.e. $Im[\chi(\Omega_{g^{'}_je_i}=0)-\chi(\Omega_{g^{'}_je_i})]$, in order to plot and compare the peak amplitudes. \\
So, by using the toy model we can easily explain the observed results analytically. From the above results we can predict that the peak amplitudes will show oscillatory behaviour for the variation of $\theta$ and $\phi$. Using these facts we can determine the magnetic field direction from the relative amplitudes of the observed peaks. In the next section we have discussed how our analytical solution can be useful in the vector magnetometry. 
\section{Detection of the unknown magnetic field }
Since the separation between the EIT peaks and the EIT peak amplitude is dependent on the orientation of the magnetic field and the polarization components of the applied electric fields, the above experimental technique can be used for the detection of the unknown magnetic field. From the separation of the two consecutive EIT resonances, we can measure the strength or the magnitude of the magnetic field. Since the peak amplitude is dependent on the quantization axis, from the relative intensity we can predict the direction of the magnetic field. 

\begin{figure}[h]
	\centering
	\includegraphics[scale=.28]{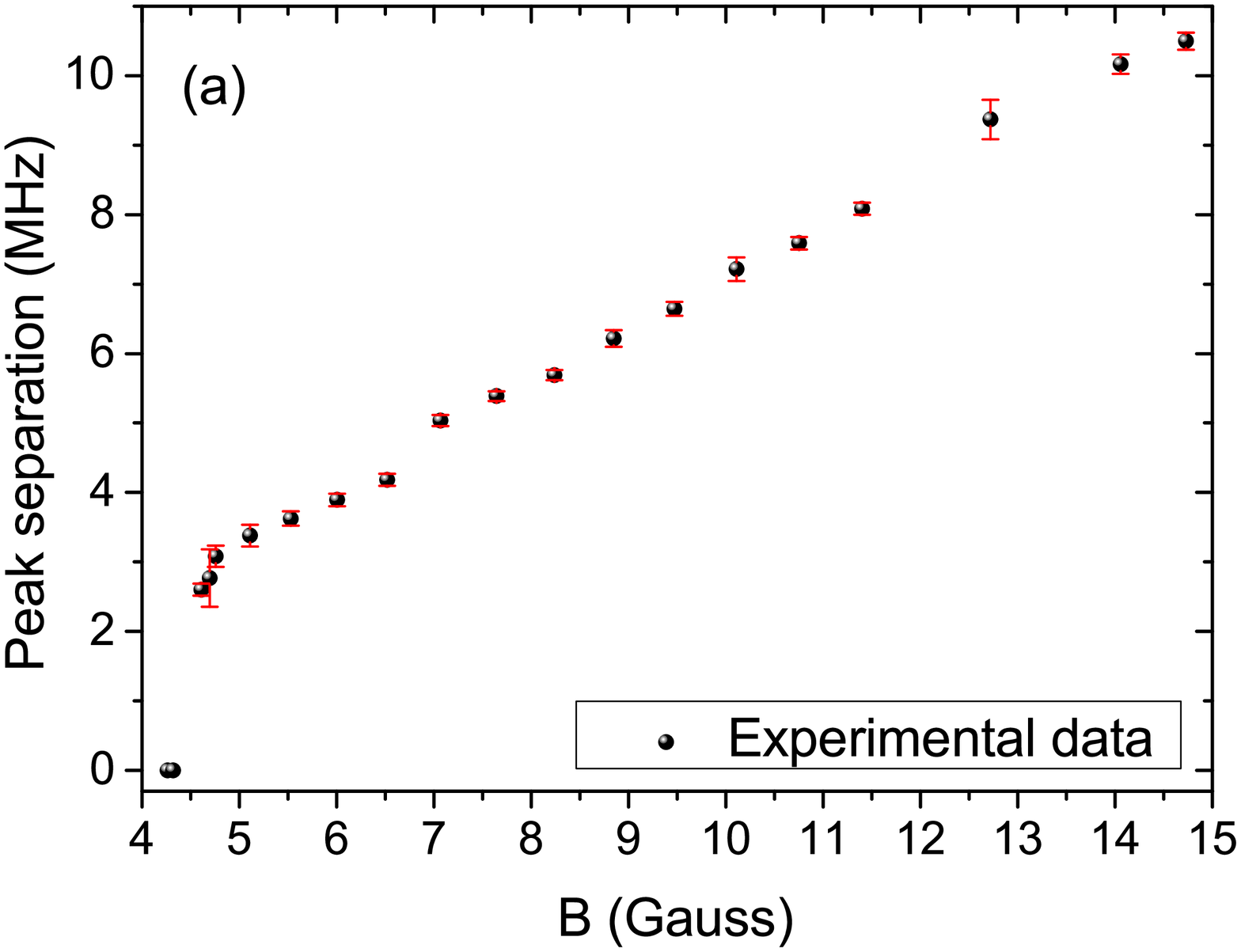}
	\includegraphics[scale=.28]{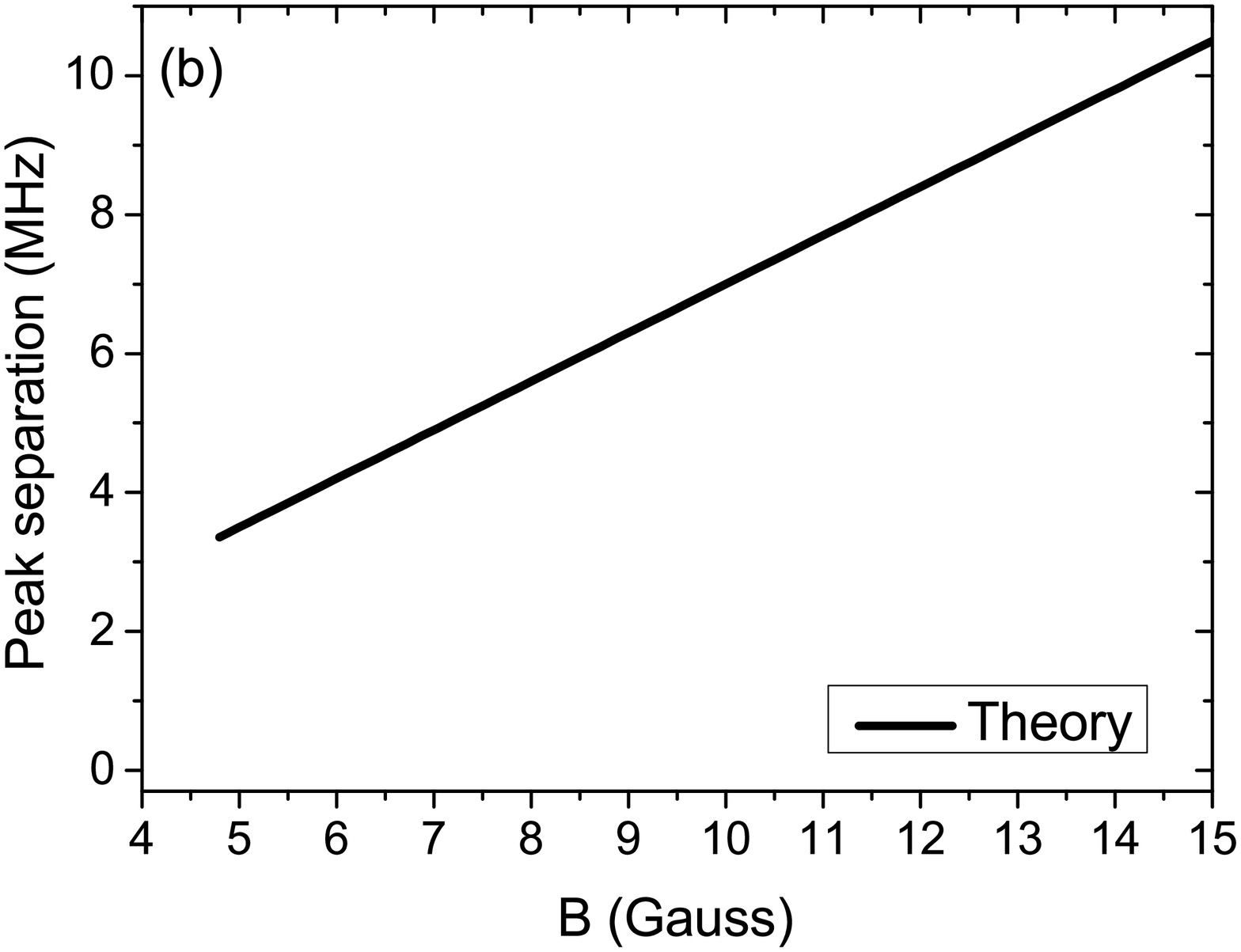}
	\caption{(a) Experimentally observed and (b) theoretically calculated, peak separations between $A_0$ and $A_{+1}$ versus magnetic field strength $|\vec{B}|$.}
	\label{magnitude}
\end{figure}

In the case of the magnitude of the field, the peak separation $\delta= \mu_B g_F |\vec{B}| \Delta m_F$, is proportional to  the magnetic field strength. For the two consecutive peaks, the peak separation becomes $\delta= \mu_B g_F |\vec{B}|$ as  $\Delta m_F =\pm 1$. In the experiment we have increased the transverse magnetic field strength ($\beta_t $) keeping the longitudinal magnetic field strength ($\beta_l $) fixed. So the total magnetic field strength will also increase since $|\vec{B}| = \sqrt{\beta_l^2 + \beta_t^2}$. In the figure \ref{magnitude}(a), experimentally observed peak separations between $A_0$ and $A_{+1}$ have been plotted with the total magnetic field $|\vec{B}|$. In the figure \ref{magnitude}(b), theoretically calculated values using $\delta= \mu_B g_F |\vec{B}|$ have been plotted. In both the cases we got similar values. We observed that when the peaks are well separated, the separation increases linearly with the increase of $|\vec{B}|$. But when the $A_{+1}$ peak just started to appear, we have observed a nonlinear behaviour in the experimentally observed values \cite{Weis2012, Weis2017}. There exists a threshold value of the magnetic field below  which the $A_{+1}$ will not be observed. In our case, it was $\beta_t=1.2$ $Gauss$ corresponding to the value of $\beta_l=4.26$ $Gauss$ (see figure \ref{magnitude}(a)). For simplicity of the calculation, in the theoretical results, we have considered only the linear part above the threshold value where the separation is proportional to the field strength.

The direction of the magnetic field can be calculated from the relative amplitudes of the $\sigma$ and the $\pi$ EIT peaks. 
 From the analytical solution of the susceptibility $\chi$, we can easily find the peak amplitudes of the individual peaks. As discussed earlier, the $A_0$ peak is coming due to the contribution of the two $\Lambda$-type systems formed by the transitions, $\ket{g^{'}_{-1}} \rightarrow \ket{e_{0}}\rightarrow \ket{g_{+1}}$ and $\ket{g^{'}_{+1}} \rightarrow \ket{e_{0}}\rightarrow \ket{g_{-1}}$. The peak amplitude can be approximately calculated as, 
\begin{equation}
\begin{array}{lll}
A_0&\simeq&Im[ \dfrac{1}{\epsilon_0 E_p}\int_{-\infty}^{+\infty}d(kv) N(kv)[\mu_{g_{+1}e_0} [ \rho_{g_{+1}e_0}(\Omega_{g^{'}_{-1}e_0}=0)-\rho_{g_{+1}e_0}(\Omega_{g^{'}_{-1}e_0})]\\
& &+ \mu_{g_{-1}e_0} [ \rho_{g_{-1}e_0}(\Omega_{g^{'}_{+1}e_0}=0)-\rho_{g_{-1}e_0}(\Omega_{g^{'}_{+1}e_0})]]]\\
&=& Im[\dfrac{i N_0 \Lambda_0}{6\epsilon_0 E_p} \dfrac{\mu_{g_{+1}e_0} \Omega_{g_{+1}e_0} |\Omega_{g^{'}_{-1}e_0}|^2}{(W_D +\gamma_{g_{+1}e_0} + i \delta)(4\gamma_{g_{+1}g^{'}_{-1}} (W_D + \gamma_{g_{+1}e_0} + i\delta)+ |\Omega_{g^{'}_{-1}e_0}|^2)}\\
&&+  \dfrac{iN_0\Lambda_0}{6\epsilon_0E_p} \dfrac{\mu_{g_{-1}e_0} \Omega_{g_{-1}e_0} |\Omega_{g^{'}_{+1}e_0}|^2}{(W_D +\gamma_{g_{-1}e_0} - i \delta)(4\gamma_{g_{-1}g^{'}_{+1}} (W_D + \gamma_{g_{-1}e_0} - i\delta)+ |\Omega_{g^{'}_{+1}e_0}|^2)}]

\end{array}
\end{equation}
In the calculation of the $A_0$ peak amplitude we considered $\Delta_p=\Delta_c=0$. Here we have subtracted the two photon contribution from the one photon Doppler contribution in order to find the peak amplitude. The term $i \delta$ has no contribution on the peak amplitude. So we have neglected it and the peak amplitude becomes,

\begin{equation}\label{A_0 peak amplitude}
\begin{array}{lll}
A_0
&=& \dfrac{N_0\Lambda_0}{6\epsilon_0E_p} \dfrac{ \mu_{g_{+1}e_0}\Omega_{g_{+1}e_0} |\Omega_{g^{'}_{-1}e_0}|^2}{(W_D +\gamma_{g_{+1}e_0} )(4\gamma_{g_{+1}g^{'}_{-1}} (W_D + \gamma_{g_{+1}e_0})+ |\Omega_{g^{'}_{-1}e_0}|^2)}\\
&&+  \dfrac{N_0\Lambda_0}{6\epsilon_0E_p} \dfrac{\mu_{g_{-1}e_0}\Omega_{g_{-1}e_0} |\Omega_{g^{'}_{+1}e_0}|^2}{(W_D +\gamma_{g_{-1}e_0} )(4\gamma_{g_{-1}g^{'}_{+1}} (W_D+\gamma_{g_{-1}e_0} )+ |\Omega_{g^{'}_{+1}e_0}|^2)}                
\end{array}
\end{equation}

In our case, the Rabi frequencies $|\Omega_{g_{+1}e_0}|^2=|\Omega_{g_{-1}e_0}|^2=\left[\dfrac{\mu_{g_{+1}e_0} E_p}{\sqrt{2} \hbar}\right]^2\left( \sin^2(\phi) + \cos^2 (\phi) \cos^2 (\theta)\right)$ and $|\Omega_{g^{'}_{-1}e_0}|^2=|\Omega_{g^{'}_{+1}e_0}|^2=\left[\dfrac{\mu_{g^{'}_{-1}e_0}  E_c}{\sqrt{2}\hbar}\right]^2\left( \cos^2(\phi) + \sin^2 (\phi) \cos^2 (\theta)\right)$. Also $\gamma_{e_ig_k}=\gamma_{e_ig{'}_j} = \Gamma/6$, where $\Gamma$ is the natural line-width and all the $\gamma_{g^{'}_{j}g_{k}}$ are assumed to be $\approx30$ $KHz$. Similarly $A_{+1}$ peak amplitude can be calculated as above. The $A_{+1}$ peak amplitude becomes, 
\begin{equation} 
\begin{array}{lll}
A_{+1}&=&Im[ \dfrac{1}{\epsilon_0 E_p}\int_{-\infty}^{+\infty}d(kv) N(kv)[\mu_{g_{0}e_{0}} \left( \rho_{g_{0}e_{0}}(\Omega_{g^{'}_{-1}e_0}=0)-\rho_{g_{0}e_{0}}(\Omega_{g^{'}_{-1}e_0})\right)+\\
& & \mu_{g_{0}e_{-1}} \left( \rho_{g_{0}e_{-1}}(\Omega_{g^{'}_{-1}e_{-1}}=0)-\rho_{g_{0}e_{-1}}(\Omega_{g^{'}_{-1}e_{-1}})\right)]]\\
\end{array}
\end{equation}
As earlier, if we drop the $i\delta$ term, the peak amplitude becomes,
\begin{equation} \label{A_1 peak amplitude}
\begin{array}{lll}
A_{+1}
&=& \dfrac{N_0\Lambda_0}{6\epsilon_0 E_p} \dfrac{ \mu_{g_{0}e_{0}}\Omega_{g_{0}e_{0}} |\Omega_{g^{'}_{-1}e_{0}}|^2}{\left(W_D +\gamma_{g_0e_0} \right)\left(4\gamma_{g_0g^{'}_{-1}} (W_D + \gamma_{g_0e_0})+ |\Omega_{g^{'}_{-1}e_0}|^2\right)}\\
&&+  \dfrac{N_0\Lambda_0}{6\epsilon_0 E_p} \dfrac{\mu_{g_{0}e_{-1}} \Omega_{g_{0}e_{-1}} |\Omega_{g^{'}_{-1}e_{-1}}|^2}{\left(W_D +\gamma_{g_0e_{-1}} \right)\left(4\gamma_{g_0g^{'}_{-1}} (W_D + \gamma_{g_{0}e_{0}} )+ |\Omega_{g^{'}_{-1}e_{-1}}|^2\right)}
\end{array}
\end{equation} 

\begin{figure}[h]
	\centering
	\includegraphics[scale=.28]{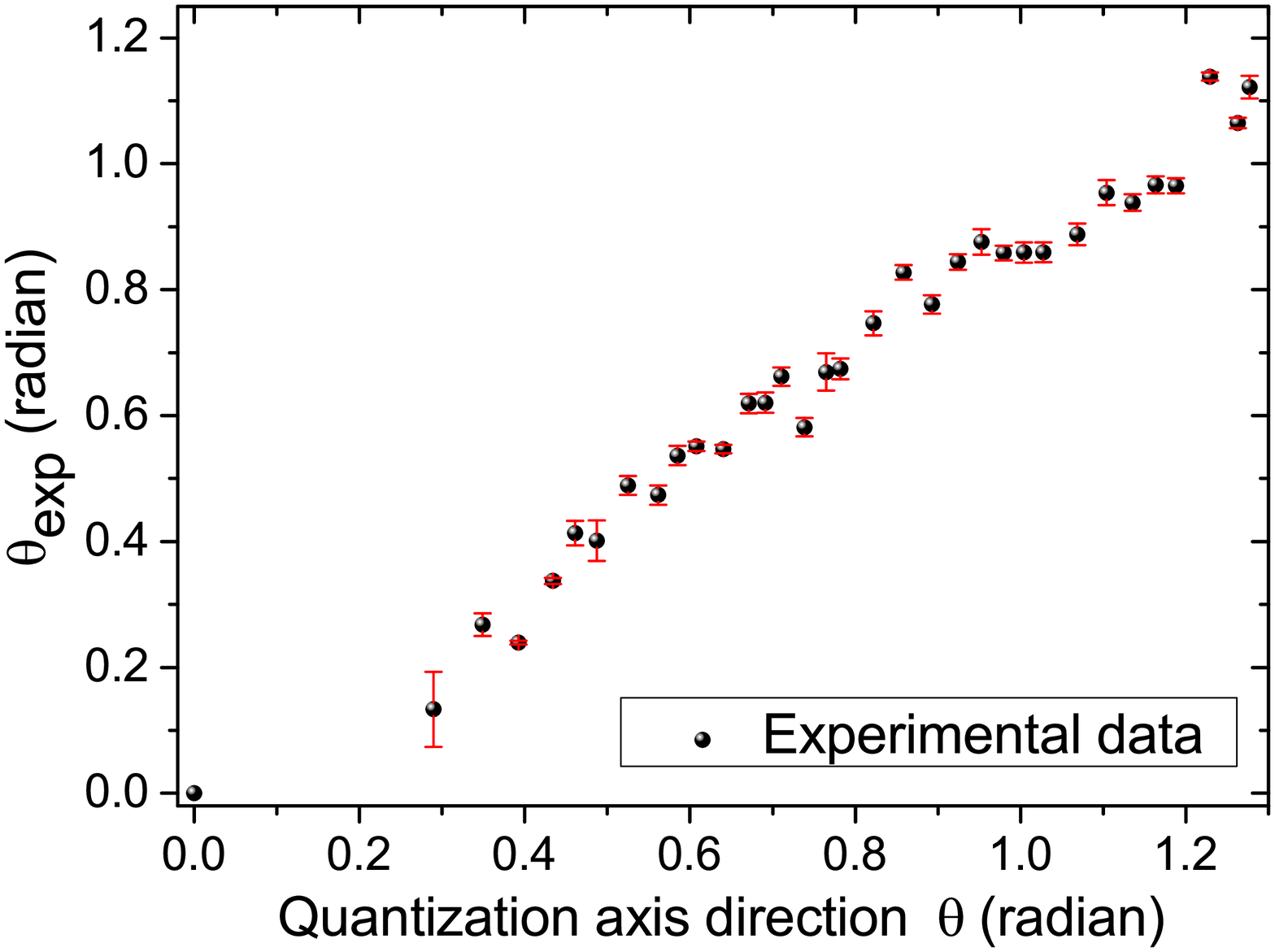}
	\includegraphics[scale=.28]{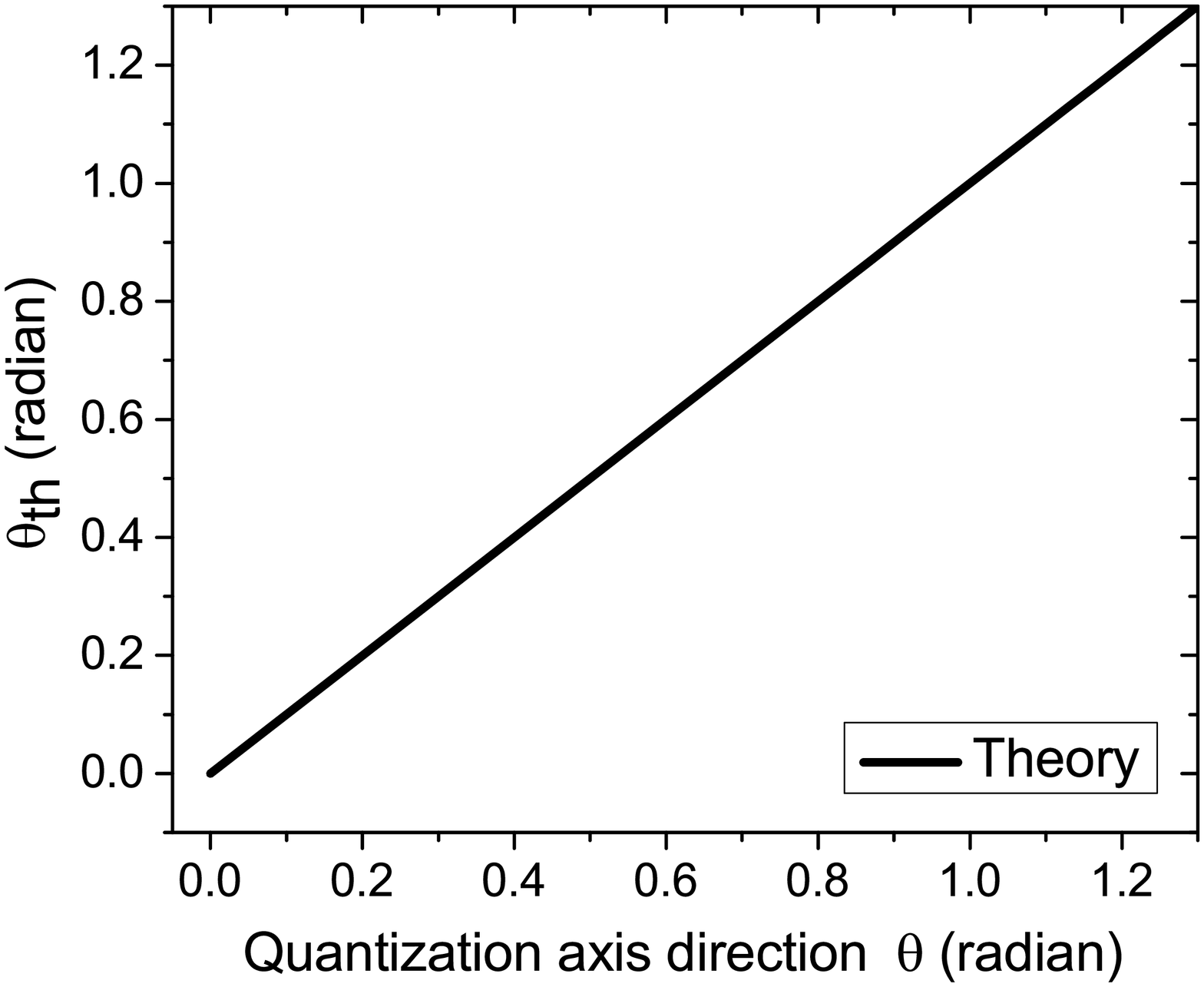}

	\caption{ Calculated magnetic field direction versus quantization axis direction ($\theta$). (a) Experimentally calculated $\theta_{exp}$ values. (b) Theoretical $\theta_{th}$ values considering equation \ref{magnetic field diretion}.}
	\label{direction}
\end{figure}

Here we assume $\Delta_p= +\delta$ and $\Delta_c=0$. The Rabi frequencies are defined as, $|\Omega_{g_{0}e_0}|^2=\left[\dfrac{\mu_{g_{0}e_0} E_p}{\hbar}\right]^2 \cos^2{\phi}\sin^2{\theta}$;  $|\Omega_{g^{'}_{-1}e_0}|^2=\left[\dfrac{\mu_{g^{'}_{-1}e_0} E_c}{\sqrt{2}\hbar}\right]^2( \cos^2(\phi) + \sin^2 (\phi) \cos^2 (\theta))$;  $|\Omega_{g_{0}e_{-1}}|^2= \left[\dfrac{\mu_{g_{0}e_{-1}} E_p}{\sqrt{2} \hbar}\right]^2( \sin^2(\phi) + \cos^2 (\phi) \cos^2 (\theta))$ and  $|\Omega_{g^{'}_{-1}e_{-1}}|^2= \left[\dfrac{\mu_{g^{'}_{-1}e_{-1}} E_c}{ \hbar}\right]^2( \sin^2(\phi) + \cos^2 (\phi) \cos^2 (\theta))$.\\

The above equations (\ref{A_0 peak amplitude} and \ref{A_1 peak amplitude}) are the theoretical formulae for calculating the peak amplitudes. Using the relative peak amplitudes we can measure the direction ($\theta$) of the magnetic field. For our specific case, we have done the experiment for $\phi =0$ and varied the direction of the quantization axis $\theta$. In this case the equation becomes even simpler if we take the ratio of the two peaks,
\begin{equation}
\dfrac{A_{+1}}{A_0}= \frac{1}{\sqrt{2}} (\frac{\mu_{g_{0}e_{0}}}{\mu_{g_{+1}e_{0}}})^2 \tan{(\theta)}
\end{equation}

So the ratio is dependent only on the direction of the magnetic field and the dipole transition strengths. Interestingly it is independent of the intensity of the laser beams i.e., it is free from power broadening effects. Using this equation we can calculate the direction of the magnetic field, 
\begin{equation}\label{magnetic field diretion}
\theta_{th}= \tan^{-1} {\left[\sqrt{2}\left(\frac{\mu_{g_{+1}e_{0}}}{\mu_{g_{0}e_{0}}}\right)^2 \dfrac{A_{+1}}{A_0}\right]}
\end{equation}

In the figure \ref{direction} we have compared the experimental results with the theoretical ones. Figure \ref{direction}(a) shows the calculated magnetic field direction $\theta_{exp}$ from the experimentally observed values. Since the intensity of the $\pi$ and the $\sigma$ EIT peaks are proportional to $\sin^2({\theta})$ and $\cos^2({\theta})$ \cite{Wynands98}, the amplitudes of $A_{+1}$ and $A_0$ were calculated by taking the square root of the experimentally observed intensities. For the theoretical calculation in figure \ref{direction}(b), we directly used the amplitudes $A_{0}$ and $A_{+1}$ from the equations \ref{A_0 peak amplitude} and \ref{A_1 peak amplitude}. The quantization axis direction $\theta$ is calculated from the magnetic field strengths using $\theta= \tan^{-1}(\frac{\beta_t}{\beta_l})$ which is an independent parameter.

So, by using the above technique we can find the unknown magnetic field strength and its direction. Therefore this formulation can be useful in developing an EIT based vector magnetometer.\\

\begin{figure}[h]
	\centering
	\includegraphics[scale=.28]{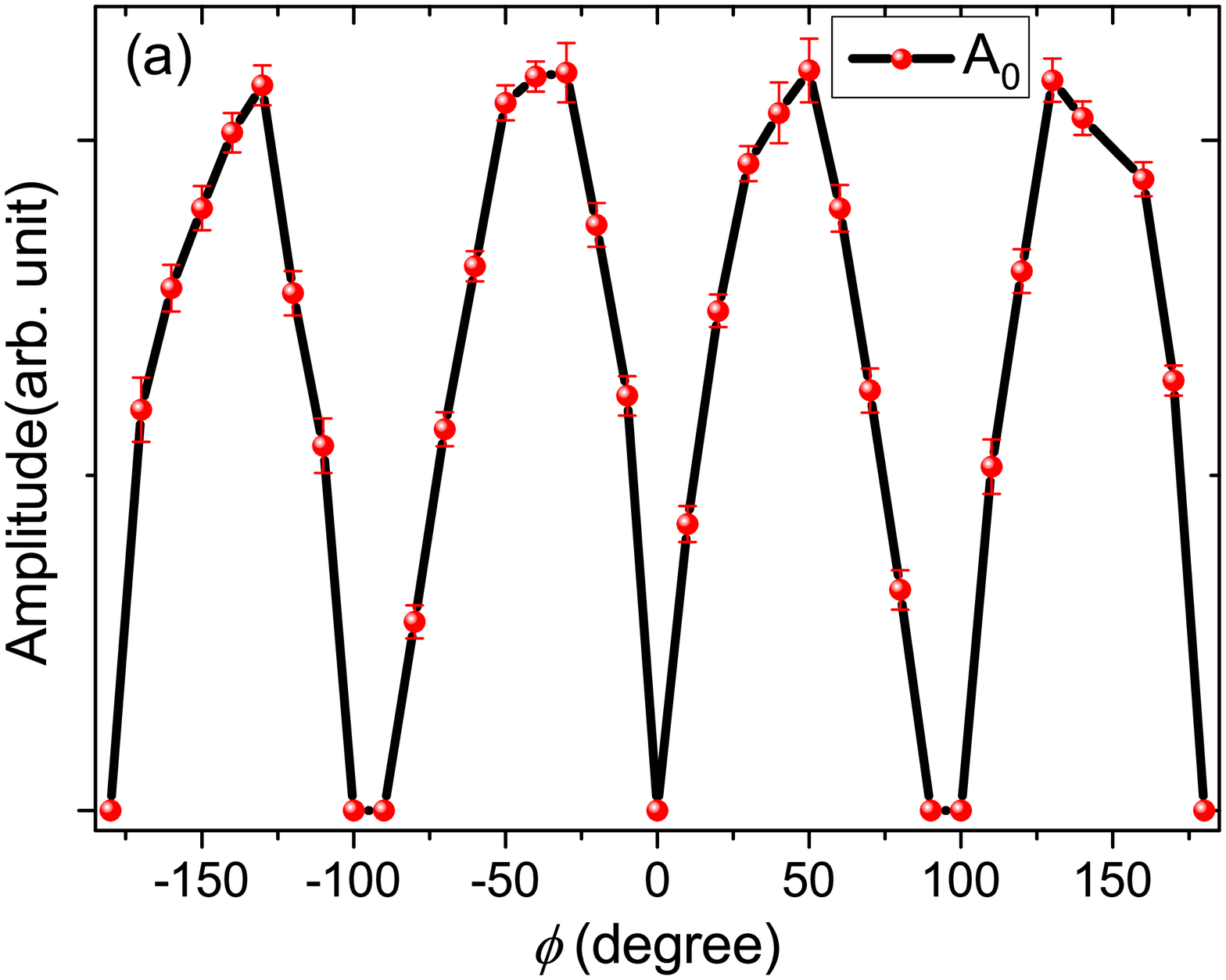}
	\includegraphics[scale=.27]{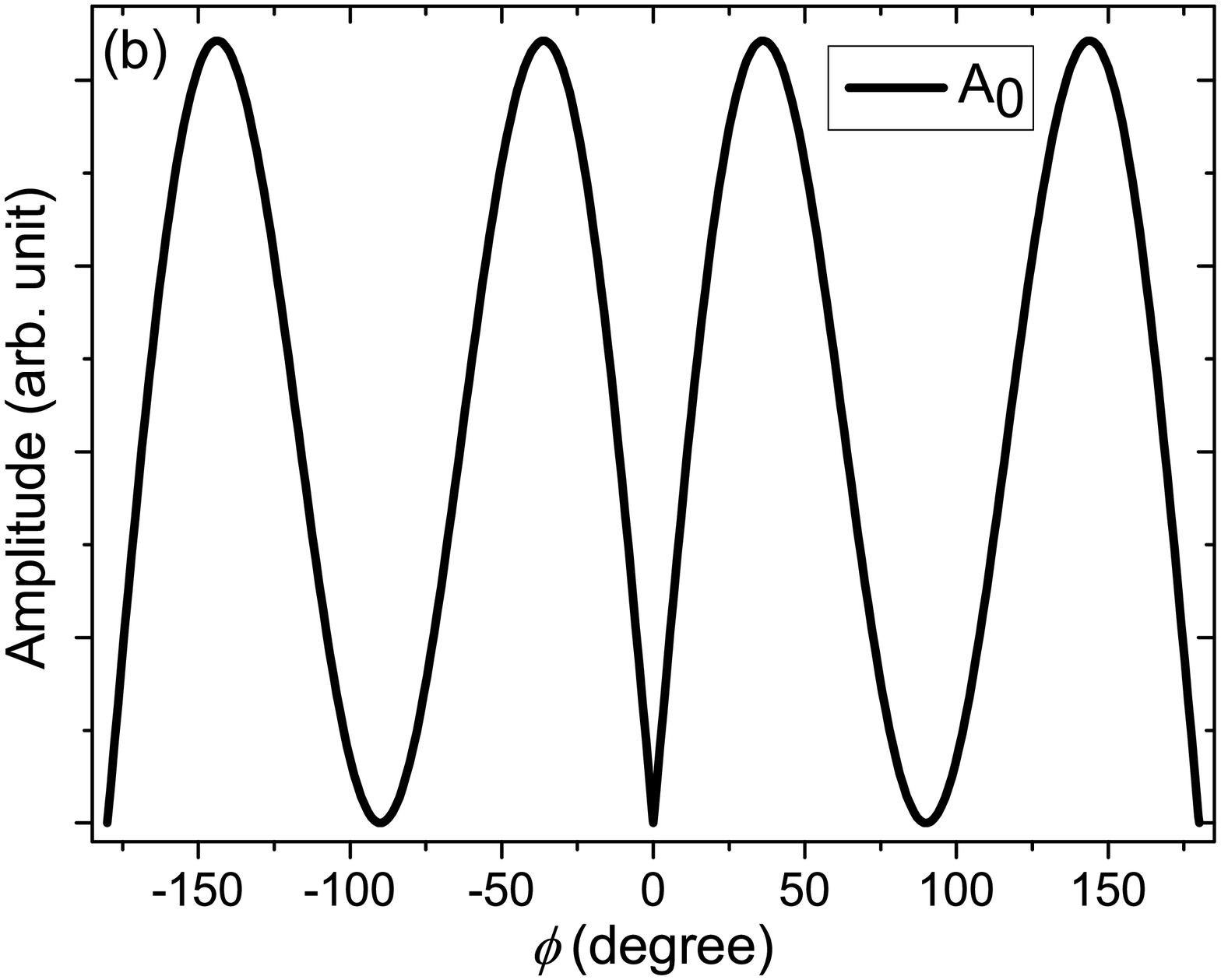}
	
	\caption{ Locus of the $A_0$ EIT peak amplitude with the variation of $\phi$ while $\theta=90^0$. (a) Experimentally observed values. (b) Theoretically plotted results with equation \ref{A_0 peak amplitude_phi}. }
	
	\label{phi_phase}
\end{figure}

Further the peak amplitudes are also dependent on the input polarization components of the electric fields. In the vector magnetometry the polarization is an important parameter for the detection of the direction of the magnetic field. We have already seen in the spectra of the figures \ref{theta_phi_variation_experiment}(b) and \ref{theta_phi_variation_theory}(b) that the observed peaks appear in an oscillatory manner. In the figure \ref{phi_phase} we have plotted the locus of the peak amplitude of $A_0$ as a function of the angle $\phi$ where the quantization axis $\theta =90^0$ is kept fixed. We observed that the peak $A_0$ shows local maxima and minima. In the equation \ref{A_0 peak amplitude} we have calculated the analytical behaviour of the $A_0$ peak amplitude as a function of $\theta$ and $ \phi$. When $ \theta =90^0$, the $A_0$ amplitude becomes,

\begin{equation}\label{A_0 peak amplitude_phi}
\begin{array}{lll}
A_0
&=& \dfrac{N_0\Lambda_0}{6\sqrt{2}\epsilon_0\hbar^3} \dfrac{ \mu_{g_{+1}e_{0}}^2 \mu_{g^{'}_{-1}e_{0}}^2 E_c^2 \sin(\phi)\cos^2(\phi) }{(W_D +\gamma_{g_{+1}e_0} )\left(4\gamma_{g_{+1}g^{'}_{-1}} (W_D + \gamma_{g_{+1}e_{0}})+  \left[\dfrac{\mu_{g^{'}_{-1}e_0} E_c}{\sqrt{2}\hbar}\right]^2 \cos^2(\phi) \right)}
\end{array}
\end{equation}

So, it is clear that when $\phi=\frac{n\pi}{2}$, where $n=1,2,3...$, the peak will not be observed as it becomes zero. 
In figure \ref{phi_phase}(a) we have plotted the experimentally observed characteristics of the peak amplitude as function of $\phi$. 
Theoretically we are getting similar results (see figure\ref{phi_phase}(b)) as observed from the experiment. The curve shows the local maxima and minima.
For the case $\phi=0$, the $\sigma$ polarization of the the probe beam is zero. So, no $\Lambda$-type system will be formed at $\Delta_p= 0$. So $A_0$ will not be observed. Similarly for $\phi=90^0$, again $\sigma$ component of the pump beam will be zero but the probe beam will have the $\sigma$ component. So at $\Delta_p=0$ there will not be any $\Lambda$-type system. So, again $A_0$ will not be observed. In between, we will get the maxima for the $A_0$ peak. Since the local minima and the maxima are dependent on the quantization direction, by finding the minima and the maxima we can find the direction of the magnetic field. If we change the direction of the magnetic field, the maxima and the minima will be shifted.
\section{conclusion}
In summery we have experimentally studied the effects of the longitudinal and the transverse magnetic fields in the EIT of a $\Lambda$-type system. We have shown how we can select the different EIT resonances by controlling the polarization components. To understand the experimental results we have calculated the probe transmission considering all the Zeeman sub-levels for a complete solution. Apart from the numerical model, we have also calculated the characteristic features of the $\sigma$ and $\pi$ EIT peaks analytically using a toy model. The EIT peaks show oscillatory behaviour with respect to the angle $\theta$ and $ \phi$. Our analytical model reveals the explicit dependency of the EIT peaks on the quantization axis direction ($\theta$) and the angle of the polarization axis ($\phi$). We have also derived the direction of the applied magnetic field ($\theta$) analytically from the relative amplitudes of the EIT peaks. Furthermore, we have shown how our analytical solution can be helpful in developing an EIT based atomic vector magnetometer. Even though we have considered a toy model, we have explained all the experimental observations and their explicit dependencies on the parameters with this simplistic model. 

Practical applications of the vector magnetometers in the field of navigation, biology, geology and industry are increasing day by day. Our analytical model can be further developed to produce precise values for the magnetic field vectors for any atomic system. These values can be compared \textit{in-situ} with the measurements from different devices. This might lead to real-time monitoring, correction and calibration of the vector fields in order to develop a realistic atomic vector magnetometer. In this way our study can be helpful in enriching the field of vector magnetometry along with the existing technologies. 

\section{Acknowledgment}

BCD, AD and SD acknowledge the financial support received from the Department of Atomic Energy (DAE), Government of India. DB thanks Science and Engineering Research Board (SERB) for Teachers Associateship for Research Excellence (TARE) award (TAR/2018/000710).

%

\end{document}